\documentstyle[psfig]{laa}
\def\solar {\ifmmode_{\mathord\odot} \else $_{\mathord\odot}$\fi}% _solar
\def\Msol {\ifmmode {\,{\rm M}\solar} \else $\,{\rm M}$\solar\fi} % solar mass
\def\Rsol {\ifmmode {\,{\rm R}\solar} \else $\,{\rm R}$\solar\fi} % solar radius
\newcommand{\less}{\raisebox{-0.6ex}{$\,\stackrel{\raisebox{-.2ex}%
{$\textstyle<$}}{\sim}\,$}}
\newcommand{\more}{\raisebox{-0.6ex}{$\,\stackrel{\raisebox{-.2ex}%
{$\textstyle>$}}{\sim}\,$}}
\begin{document}
\thesaurus{
              06         % A&A Section 6: Form. struct. and evolut. of stars
% List of keywords in A&A 305 #2
              (
               08.02.3;  % Stars: binaries, general
               08.02.4;  % Stars: binaries, spectroscopic
               08.02.6;  % Stars: binaries, visual
               08.12.2)  % Stars: low-mass, brown dwarfs.
     } 
%
% 13--> 11 Enleve la composante externe de LP476-207, vue par Henry et al. 97,
% et G203-47 vue par Reid &Gizis
% 11--> 12 Ajoute GJ 2130A
% within 9 parsecs
% Commente pour le moment car plus vrai... Eventuellement reflechir a un
% titre plus accrocheur 
\title{
New neighbours. I. 13 new companions to nearby M dwarfs\thanks{Partly based on
observations made at Observatoire de Haute-Provence,
operated by the Centre National de la Recherche Scientifique de France and on 
observations made at Canada-France-Hawaii Telescope, 
operated by the National Research
Council of Canada, the Centre National de la Recherche Scientifique de
France and the University of Hawaii.}}
%
%\subtitle{ }
\author{X. Delfosse \inst{1,2}, T. Forveille \inst{1}, J.-L.
          Beuzit \inst{2,3}, S. Udry \inst{2}, M. 
          Mayor \inst{2}, C. Perrier \inst{1} 
       }
\offprints{Xavier Delfosse, e-mail: Xavier.Delfosse@obs.ujf-grenoble.fr}
\institute{Observatoire de Grenoble,
              Universit\'e J. Fourier, BP53,
              F-38041 Grenoble,
              France
\and
              Observatoire de Gen\`eve,
              CH-1290 Sauverny,
              Switzerland
\and
              Canada-France-Hawaii Telescope Corporation, 
              P.O. Box 1597,
              Kamuela, HI 96743, 
              U.S.A.
             }
\date{Received ; accepted }
\maketitle
\markboth{Delfosse et al. }{Thirteen new nearby M dwarfs} 
\begin{abstract}
We present preliminary results of a long-term radial-velocity search
for companions to nearby M dwarfs, started in September 95. The
observed sample is volume-limited, and defined by the 127 northern
(${\delta}>-16^{\circ}$) M
dwarfs listed in the Gliese and Jahreiss (CNS3) catalogue with d$\le$9pc and
V$\le$15. Observations are obtained with the ELODIE spectrograph on the
1.93-m telescope of the Observatoire de Haute-Provence. The typical
accuracy ranges between 10~$\rm{m\,s^{-1}}$ (the instrumental stability
limit) for the brighter stars and 70~$\rm{m\,s^{-1}}$ at our limiting
magnitude. We complement the ELODIE velocities with older measurements
extracted from the CORAVEL database to extend our time base, albeit
obviously with lower precision.  Simultaneously, we perform adaptive
optics imaging at CFHT and ESO to look for close ($a>$0.05-0.1'') visual
companions in a larger volume-limited sample. For stellar companions the two
techniques together cover the full separation range, to beyond the
limiting distance of the sample. We will therefore eventually obtain a
statistically meaningful inventory of the stellar multiplicity of nearby
M-dwarf systems.
We also have useful sensitivity to giant planets, as illustrated by
our recent detection of a planetary companion to Gl~876.

% Attention, GJ2130 est a mettre bien a part dans toute discussion 
% statistique. Donc 12, pas 13...
After 2.5 years, we have discovered 12 previously unknown components in
this 127 stars sample, plus a companion to
an additional star beyond its southern declination limit. 7 of these 
are actually beyond the 9~pc limit, as they belong to systems included in the 
sample on the basis of CNS3 photometric parallaxes which were biased-down 
by the unrecognized companion. The remaining 5 companions are true additions
to the 9~pc inventory. More are certainly forthcoming, given our present
selection bias towards short periods and relatively massive 
companions.

% Xavier, je ne comprend pas pourquoi le paragraphe ci-dessous ne te convenait
%  pas. Dans le doute j'ai remis en anglais ton texte modifie plutot que de 
% restaurer tel quel mon paragraphe, mais si tu fais d'autre modifs de ce 
% genre met une note explicative SVP.
% Thierry
%
%A number of the new
%multiple systems had no trigonometric parallax in the CNS3, and were
%initially included in the nominal d$<$9pc volume on the basis of a
%CNS3 photometric parallax. Correcting the photometric distance for
%multiplicity, 7 of these are actually beyond the 9~pc limit, as
%confirmed for most by recent HIPPARCOS or ground-based trigonometric
%parallaxes. New distances exclude another 6~systems from the 9~pc
%volume, and add only one new system. These changes increase the
%significance of the known incompleteness of the nearby M dwarf sample
%beyond 5~pc. The new companions on the other hand add 5 stars to the
%sample volume, with a strong selection bias towards massive companions
%at short periods. This indicates that it is still premature to derive
%a low-mass star luminosity function from local samples beyond
%$\sim$5~pc.

We have derived orbital elements for 7 of the new systems, as well
as for some known binaries with previously undetermined orbits. One
system, G~203-47, associates an M3.5V star with a white dwarf in a
rather tight orbit ($a_1\sin{i}~=~15\Rsol$) and represents
a Post-Common-Envelope system. Some of
the new M-dwarf binaries will over the next few years provide very precise
mass determinations, and will thus better constrain the still poorly determined
lower main-sequence mass-luminosity relation. The first such results are
now being obtained, with some preliminary accuracies that range between
2\% at 0.4-0.6~{\Msol}  and 10\% at 0.1~{\Msol}. We have also discovered the 
third known detached M-dwarf eclipsing binary, and determined its 
masses with 0.4\% accuracy.

  \keywords{stars: binaries - stars: low mass, brown dwarfs - planets: giant
  planets - techniques: radial velocity - techniques: adaptive optics}

\end{abstract}

\section{Introduction}

Binarity is a key observational parameter for many astrophysical
questions. One particularly important issue is whether low mass
stars and brown dwarfs significantly contribute to the dark 
matter in the galactic disk; the present uncertainties on the local mass
function for very low mass stars are large enough for this question to be 
still unsettled (Tinney 1993), although evidence increasingly points towards 
a negative answer.
% if the IMF cuts-off at the hydrogen burning limit. 
%% Clairement non: Delfosse et al. 1997...
The DENIS and 2MASS near infrared surveys will soon provide
photometric luminosity functions with very good statistical precision,
down to the brown-dwarf domain (Delfosse et al. 1997, 1998a;
Kirkpatrick et al. 1997). The corrections for unresolved binaries and
the mass-luminosity relation will then be the main uncertainty
sources for the local stellar mass function. Below M$\sim$0.3~{\Msol}, 
the mass-luminosity relation (Henry \& McCarthy 1993) is
only determined by few observational data points, most of which
are of low accuracy. Accurate orbital elements are thus clearly needed for more
low mass binaries, to provide additional direct stellar mass determinations.
A better knowledge of the multiplicity statistics of low
mass stars is also needed to account for the effect of unresolved systems on 
the mass and
luminosity functions: at present, different plausible assumptions on
stellar multiplicity (Kroupa 1995; Reid \& Gizis 1997) lead to very
different luminosity functions at very low masses.
The stellar multiplicity statistics also contains important information
on both star formation processes and dynamical evolution of stellar 
systems, as reviewed for instance by Duquennoy \& Mayor (1991). 
%% Ne me parait pas coller avec la logique generale de l'introduction.
% The search for very low mass stars and brown
% dwarfs is also historically a strong motivation for the multiples
% studies, the best statistic place to discovery a stellar or substellar
% objects being near a star. The first undisputed field brown dwarf 
% Gl~229B (Nakajima et al. 1995) and the brown dwarf candidates GD165B
% (Becklin \& Zuckerman 1988) were both found as companions to brighter
% stars.

While the binarity statistics are now quite well determined for G
(Duquennoy \& Mayor 1991) and K dwarfs (Halbwachs, Mayor and Udry
1998), this is not yet the case for the fainter M~dwarfs. The only
well defined M-dwarf sample which has essentially complete
multiplicity information is the (small) sample of M dwarfs within
5.2~pc (Henry \& McCarthy, 1990; Leinert et al., 1997). A number of
programmes have searched for M-dwarf companions beyond this distance,
using different techniques: radial-velocity monitoring
% with a typical accuracy of 200~$m\,s^{-1}$, and many with a typical 
% accuracy of 1--2~$km\,s^{-1}$ 
(Bopp \& Meredith 1986; Young et al. 1987; Upgren \& Caruso 1988;
Marcy \& Benitz 1989; Tokovinin 1992; Upgren \& Harlow 1996), deep
infrared imaging (Skrutskie et al. 1989; Nakajima et al. 1994; Simons
et al. 1996), astrometry
%(ref??), 
and speckle or adaptive optics imaging (Henry \& McCarthy 1992;
Mariotti et al. 1992).  Taken together, these programmes are however
not sensitive to all binary separations for any statistically well
defined sample.  Reid \& Gizis (1997) for instance compiled
multiplicity information for a northern 8-pc sample, but had to
conclude that, in spite of Henry \& McCarthy's (1992) extensive
speckle work, the information they could gather remained incomplete.
%XD: APRES RETOUR DU REFEREE
%Nevertheless, a few attemp exist in the litterature, to determine the binarity
%fraction of the M dwarfs. They are based either about a limited sample or 
%either in considering a very important correction of completudes. In the 
%first categories we can found the work of Leinert et al. (1997,
%binarity fraction of 26$\pm$9\%) or that of Reid and Gizis (1997,
%binarity fraction of 33\%) based respectively about the 5.2~pc and
%8~pc sample. But the first sample is very small and has a poor
%statistic and the second is already incomplete. The work of Fisher and
%Marcy (1992, binarity fraction of 42$\pm$9\%) belong to the second
%categories, the main difficulty to give a important correction of
%completudes is than they are strongly dependant of the multiplicity
%statistics (mass ratio, period distribution. etc.) wich are unknow for
%this stars. All this work point out a binarity fraction less important
%for the M dwarfs than for the G dwarfs, but a definitive confirmation,
%and a accurate determination, is already to do. It is now primordial
%to make a searching be able to find all the nearby M~dwarfs companions
%in a well defined volume-limited sample.
% TF, rendu plus concis.
The binary fraction derived from these data range from 26\% (Leinert
et al., 1997) to 42\% (Fisher and Marcy, 1992), through 33\% (Reid and
Gizis, 1997). This seems to points towards a smaller fraction of
multiple stars than the 57\% found by Duquennoy \& Mayor (1991)
amongst G dwarfs, a result which would be an important input to
multiple star formation theories. All of these estimates however
suffer from either small number statistics or large incompleteness
corrections, which are quite uncertain, as they sensitively depend on
the (unknown) underlying multiplicity distribution.

Since September 1995, we have therefore been searching volume-limited
samples of nearby M~dwarfs for companions, combining three observing
techniques which together ensure a good sensitivity at all separations
for stars and brown dwarfs, and some useful sensitivity for giant
planets:
\begin{itemize}
\item{} highly accurate (10-70~${\rm m\,s^{-1}}$) radial velocities are
measured
with the ELODIE spectrograph at the Observatoire de Haute-Provence
(Baranne et al. 1996), which was also used in the discovery of the
first extra-solar planet (Mayor and Queloz 1995); they are combined
with older CORAVEL data to extend our time base at lower accuracy;
\item{} high angular resolution near infrared images are obtained with the ESO
(ADONIS) and CFHT (PUE'O) adaptive optics systems, to directly detect
relatively massive companions at small separations;
%XD APRES RETOUR DU REFEREE: precision du status de la recherche
%coronographique. TF: deplace ici
\item{} deep adaptive optics coronographic infrared images (Beuzit et
al. 1997) are obtained with ADONIS at ESO, since April 1997; they are
sensitive to lower mass companions at slightly larger separations, and
will be discussed in a forthcoming paper.
\end{itemize}

In this first paper, we present the radial-velocity programme, discuss
its observing and data processing techniques in some detail, and
present its first results together with some complementary adaptive
optics data obtained with PUE'O.
The observed sample and its completeness are discussed
in Sect. 2 together with the observing strategy, while Sect. 3 discusses 
the radial-velocity and 
adaptive optics observations. Section 4 presents the results after
2.5~years, namely the discovery of 13 new M dwarfs and orbital elements
for 7 of the new systems.  Section 5 discusses the implications of the new
systems for the solar neighbourhood population.

\section{Sample selection and observing strategy}

\subsection{The sample}
The observed radial-velocity sample is presented in detail by Delfosse
et al. (1998b), who discuss the rotational properties of the same
stars. Briefly, it contains all M dwarfs listed in the third edition
of the nearby star catalogue (hereafter CNS3; preliminary version, Gliese \&
Jahreiss, 1991) with a distance closer than 9~pc and a B1950.0
declination above $-16^{\circ}$. 136 stars fulfill these criteria
among which 7 have to be rejected because they are fainter than V=15 (the
approximate sensitivity limit of the instrument we use), and another 2 because
they are close companions to much brighter G dwarfs from which they cannot be
separated by the spectrograph input fiber. At 9~pc the V=15 limit corresponds 
to an M6 dwarf, and at 5~pc to an M6.5 dwarf.

We have decided to observe a volume-limited sample because we are
interested in a fair sampling of the local dwarf population, and want
to derive unbiased statistical information.  The quality of the
distances listed in the CNS3 is however uneven, so this goal could
only be partly reached. The new (4$^{\rm th}$) edition of the Yale General
Catalogue of Stellar Trigonometric Parallaxes (Van Altena et al. 1995)
and the HIPPARCOS catalogue (ESA, 1997) together show that 11 sample
stars are actually outside the nominal 9-pc sphere, while two omitted
stars are actually within it. 
%paragraphe retire de l'abstract
%XD: A mon avis cela ne colle pas tres bien ici, mais va plutot dans les 
%conclusions...
%A number of the new multiple systems had no trigonometric parallax in the
%CNS3, and were
%initially included in the nominal d$<$9pc volume on the basis of a
%CNS3 photometric parallax. Correcting the photometric distance for
%multiplicity, 7 of these are actually beyond the 9~pc limit, as
%confirmed for most by recent HIPPARCOS or ground-based trigonometric
%parallaxes. New distances exclude another 6~systems from the 9~pc
%volume, and add only one new system. These changes increase the
%significance of the known incompleteness of the nearby M dwarf sample
%beyond 5~pc. The new companions on the other hand add 5 stars to the
%sample volume, with a strong selection bias towards massive companions
%at short periods. This indicates that it is still premature to derive
%a low-mass star luminosity function from local samples beyond
%$\sim$5~pc.
%fin de paragraphe de l'abstract
The 11 rejected
stars have a disproportionately large fraction of nearly equal-mass
multiple systems. Most of them were included in the initial sample
on the basis of CNS3 photometric parallaxes.  
% Pas toutes: Gl2066, Gl49... Ces dernieres restent cependant peu au
% dela de 9pc, contrairement aux autres qui sont loin.
Their inclusion therefore largely reflects the 2$^{3/2}$ volume
bias incurred in using photometric parallaxes for unrecognized equal-mass
binaries.

%{\em 9 etoiles n'ont encore que des parallaxes photometriques, dont 3 a 
%exclure sur cette base}

\subsection{Observing strategy}

\begin{figure*}
\psfig{height=12cm,file=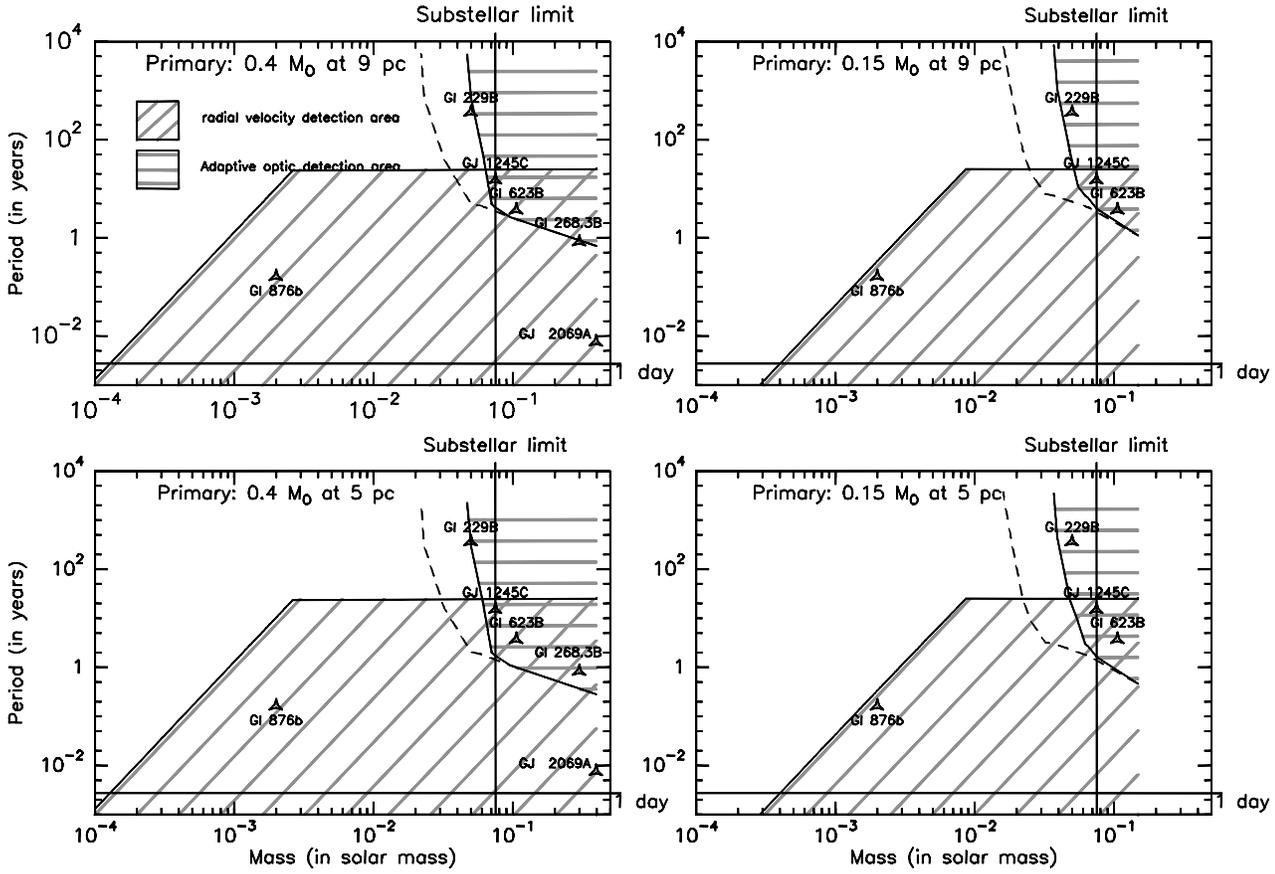,angle=-90}
\caption{Companion detectability in the $\log(P)$ versus $\log(M)$ plane 
for our radial-velocity and adaptive optics programmes. The 4 diagrams 
correspond to 
two representative primary masses of 0.4{\Msol} ($\sim$~M2V, $M_K=6.3$, 
Baraffe et al. 1998) and 0.15{\Msol} ($\sim$~M5V, $M_K=8.3$, Baraffe et al. 
1998), at two representative distances of 5 and 9~pc. 
Typical radial-velocity accuracies are respectively 12 and 70~$\rm{m\,s^{-1}}$ 
for 0.4{\Msol} and 0.15{\Msol} stars. The radial-velocity detectability limits
correspond to minimum amplitudes of 4$\sigma$, or respectively 
50 and 300~$\rm{m\,s^{-1}}$, and a maximum period of 25~years.
Some longer period binaries will be detected, but the observing
programme is unlikely to last long enough to determine their
orbital elements.
The solar metallicity models of the E.N.S Lyon group (Allard et al. 1996; 
Baraffe et al. 1998; Allard, Baraffe and Chabrier, private communication) 
for ages of 1 and 5 Gyr (dashed and solid lines, respectively) are 
used to transform the maximum detectable luminosity contrasts for PUE'O 
into approximate minimum companion masses. 
% Ne me parait pas necessaire ici: on a simplement utilise les modeles ABC 
% les plus apropries dans chaque cas, je ne crois pas necessaire d'expliquer
% le pourquoi du comment.
% For a stellar companion we consider the no-dusty NextGen model
% (Allard et al. 1995, Baraffe et al. 1998), and for a brown dwarfs companion 
% the COND model (Lyon group, private communication) wich include mollecular
% condensation equilibra but not dust opacities (i.e. that modelize a complete 
% settling).
A few representative companions of M dwarfs have been added for illustration.
Gl~623B and GJ~1245C are two of the least massive objects with dynamical mass 
determinations (Henry and McCarthy 1993). Gl~268.3B and GJ~2069Ab were 
discovered in the present programme. The brown dwarf Gl~229B (Nakajima et
al. 1995) has a most probable mass of 0.05{\Msol} (Allard et al. 1996). Gl~876b
is the first planet discovered around an M dwarf (Delfosse et al. 1998d).}
\label{limit}
\end{figure*}

Since no single observing technique is sensitive to companions over
all separations, we have chosen to combine imaging and radial-velocity
techniques, to maximize our sensitivity range. 

As discussed below, the standard errors of the radial-velocity measurements
range between 10 and 70~$\rm{m\,s^{-1}}$, depending on apparent magnitude 
(and hence largely on spectral type, for this volume-limited sample). A 
400~$\rm{m\,s^{-1}}$ velocity amplitude is thus always easily detectable, 
even for the faintest stars. A 50~$\rm{m\,s^{-1}}$ velocity amplitude is 
easily detectable for all stars brighter than approximately V=10. 
Figure \ref{limit} shows how these conservative minimum velocity 
amplitudes translate into minimum detectable companion masses as a function
of orbital period, for two representative distances and primary masses.
For a one year orbital period, a 400~$\rm{m\,s^{-1}}$ velocity amplitude 
corresponds to a 5~Jupiter mass (M$_{\rm J}$) body orbiting a 0.15~{\Msol} 
M4V-M5V primary, and 50~$\rm{m\,s^{-1}}$ corresponds to a 1.5~M$_{\rm J}$ body 
orbiting a 0.4~{\Msol} M2V primary. Even though this programme was not chiefly 
devised to search for planets, it therefore has good sensitivity to them, as 
illustrated by our recent detection of a $\sim$2~{M$_{\rm J}$} companion on
a 2~month orbit around Gl~876
(Delfosse et al. 1998d; also Marcy et al. 1998). We will, a fortiori,
be sensitive to all stellar and substellar companions with periods up
to a few dozen years. The maximum period at which we can actually detect
stellar companions will be set by the duration of the observing 
programme rather than by the velocity amplitude (Fig. \ref{limit}), as 
it is difficult to detect variability on timescales longer than about 
twice the time-span of the data. At 9~pc a limit of P${\sim}20$~years 
translates into a minimum angular separation of ${\sim}$0.4'' for a 
0.15~{\Msol} primary star and a companion of negligible mass 
(Fig.~\ref{limit}; this separation scales as $(M_1+M_2)^{1/3}$).

The adaptive optics programme (hereafter AO) (Mariotti et al. 1992,
Perrier et al. 1998) easily resolves equal-mass binaries down to
separations as small as $0.08''$ (0.7~AU at 9~pc, corresponding to a
$\sim$1~year period for an M-dwarf pair). It has a dynamic range
in the K band (${\lambda}_0~=~2.2~{\mu}$m) of 6.5~mag at a 0.25" separation,
7.5~mag at 0.5'' and 8.5~mag at 1''. Since the K band absolute magnitudes 
of M0 and M8 dwarfs only differ by $\sim$5~mag (Leggett 1992), the AO 
observations can detect all main-sequence companions that the radial-velocity
monitoring would miss (Fig. \ref{limit}). For brown dwarfs on the other hand,
we do have a small sensivity gap at intermediate separations: radial-velocity
monitoring can for instance only detect a 0.05{\Msol} Gl~229B-like brown
dwarf orbiting a 0.4{\Msol} M2V primary out to 5.4~AU (0.6'' at 9~pc), 
while AO imaging can only detect it beyond 1''. This gap disappears for 
later M-dwarf primaries or more nearby systems (Fig. \ref{limit}), and
the PUE'O adaptive optics system for instance easily detects Gl~229B 
itself, without a coronograph. Beyond 2'' (18~AU at 9~pc), the
coronographic mode of the ESO adaptive optics system (Beuzit et
al. 1997) can handle an extremely large contrast between the primary and a
faint secondary (${\Delta}m{\sim}12.5~{\rm mag}$ at 2''). We are thus again 
sensitive to all brown dwarf companions of M dwarfs, though not to planets.

Within the [0.4\,AU, 4\,AU] separation range (at 9~pc), the
sensitivities of high precision radial velocities and adaptive optics
overlap. In this range, the combination of the two techniques offers
potential access to very accurate masses, as exemplified by Gl~570B
(M1.5V+M3V) for which Forveille et al. (1998) obtain masses accurate
to $\sim$2\%. This should dramatically improve the mass-luminosity
relation for the bottom of the main sequence.

\section{Observations}

\subsection{ELODIE radial-velocity data}

\subsubsection{Instrumental setup}
Most new radial-velocity observations were obtained at the Observatoire de
Haute-Provence with the ELODIE spectrograph (Baranne et al.  1996) on
the 1.93-m telescope, between September 1995 and May 1998. This fixed
configuration dual-fiber-fed echelle spectrograph covers in a single
exposure the 390-680~nm spectral range, at an average resolving power
of 42000.  An elaborate on-line processing is integrated with the
spectrograph control software, and automatically produces optimally
extracted and wavelength calibrated spectra, with algorithms described
in Baranne et al. (1996). All stars in this programme are now observed with
a thorium lamp illuminating the monitoring fiber, as needed for the best
($\sim 10~{\rm m\,s^{-1}}$) radial velocity precision (Baranne et al. 1996).
Before mid-1997
% ?? Ben oui, apres verification c'est plus recent que je ne croyais, 
% 1245 etait encore observe en OBJ2 en juillet 1997...
the fainter (V{\more}13~mag) stars were instead observed with this fiber
illuminated by the sky, allowing subtraction of the diffused solar
light whose lines could otherwise bias the velocity profile. The use
of the new M4V correlation template (discussed below) has sufficiently
reduced the sensitivity of the velocity measurements to sky emission
that the standard high precision setup can now be used for all stars
in this programme. % Also made it necessary, by improving S/N
These early data for faint stars have increased random zero-point errors,
up to $\sim 100~{\rm m\,s^{-1}}$.

A few measurements were also obtained with the CORALIE spectrograph on the
newly commissioned 1.2-m Swiss telescope at La Silla Observatory
(Chile). CORALIE is an improved copy of ELODIE and has very similar
characteristics, with the exception of a substantially improved intrinsic
stability and higher spectral resolution ($R=50000$). 

\subsubsection{Data processing}
The extracted spectra are analysed for velocity by digital
cross-correlation with a one-bit (0/1) template. This processing is
standard for ELODIE spectra (Queloz 1995a, 1995b), but the default
K0III mask provided in the ELODIE reduction software is a relatively
poor match to the much redder spectra of the programme stars. We have
constructed a better adapted correlation template, applying the
method of Baranne, Mayor and Poncet (1979) to a high signal-to-noise
(S/N=70) ELODIE spectrum of Gl~699 (Barnard's stars, M4V). The analysis
was restricted to the [443\,nm,~680\,nm] spectral range, 
% Commente car ce n'est pas l'ordre d'interference mais juste une numerotation
% interne a INTERTACOS. Je ne connais pas la correspondance, et l'information
% ne me parait en tout etat de cause pas tres utile.
% (20$^{\rm{th}}$ to 67$^{\rm{th}}$ ELODIE orders), 
since bluer orders contain very little flux,
for both Gl~699 and the programme stars. For each of the 48 ELODIE
orders, 100 trial templates $g_0(\lambda)$ were generated by
thresholding the rectified and normalized Gl~699 spectrum at 100
levels. We selected for each order the thresholding level which
maximizes the quality factor defined by Baranne, Mayor and Poncet
(1979). The best order templates were then assembled into one global
binary template, which has over 3100 ``transparent'' sections and a
high overall transmission of 20.75\%.

\subsubsection{Measurement accuracy}

\begin{figure*}
\psfig{height=12cm,file=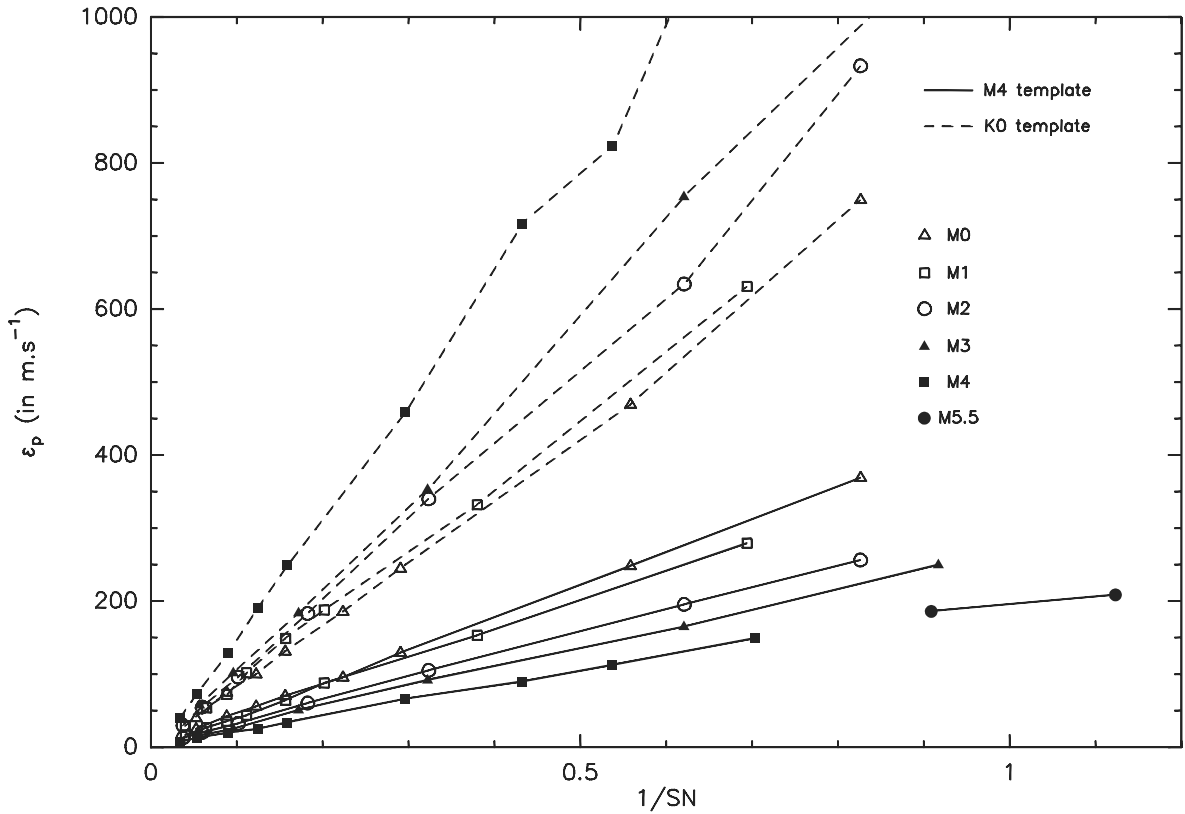,angle=-90}
\caption{Internal standard errors on radial velocity as a function of 
signal-to-noise ratio, for the two correlation templates (M4 and K0), and for
stellar
spectral types between M0 and M5.5.}
\label{error}
\end{figure*}

For ELODIE, the radial-velocity precision can be written (Baranne et
al. 1996, their Eq. (9)) as:
\begin{displaymath}
\varepsilon_p(V_r)=\frac{C(T_{eff})}{D\,S/N}\,\frac{(1+0.47{\sigma})}{3}\,{\rm{km\,s}^{-1}}
\end{displaymath}
%{\em Pourquoi le 3 n'est-il pas rentre dans C(Teff)??}\\
where ${\sigma}$ is the % (noiseless) 
rotationally broadened 1/e half-width of the cross-correlation function
(CCF), $D$ its relative depth and $S/N$ the signal-to-noise ratio in
the reference 48$^{\rm{th}}$ ELODIE order. The $C(T_{eff})$ constant
depends on both the correlation template used and the spectral type.
Its values for the K0 and M4 templates were determined through Monte-Carlo 
simulations for the [M0V,~M5.5V] spectral-type range. High signal-to-noise
ratio spectra of 6 slowly rotating stars spanning the desired range of
spectral type
% (Gl~424, Gl~15A, Gl~411, Gl~725A, Gl~699 and GJ~1002, 
(selected from Delfosse et al. 1998b, Table~\ref{caract}) 
were degraded by synthetic photon and readout noise. The dispersion 
of the velocities measured from these spectra was then determined for a number
of degraded signal-to-noise ratio values.
% (equivalent of the CCD use with ELODIE wich have 10e of noise) 
The validity of the parameterization was verified on those synthetic data,
and $C(T_{eff})$ was determined through a least square fit. It turned
out to be constant over the M-dwarf spectral range. Its respective
%XD APRES RETOUR DU REFEREE: change les unites de C pour etre coherent
%avec les unites de la formule.
values for the K0 (Baranne et al. 1996) and M4 templates are 0.085 and 
0.035~${\rm km\,s^{-1}}$. The radial-velocity standard errors 
for the two templates are presented in Fig. \ref{error}. The M4 template 
improves the accuracy by a factor of 2 at spectral type M0V, and by a
factor of 10 at spectral type M4V. At a given signal-to-noise ratio,
the accuracy variations as a function of spectral type mostly reflect
changes in the relative depth of the CCF. The width of the CCF 
(Table \ref{caract}) is essentially constant for the K0 template but
significantly varies with spectral type for the M4 template. This
change is small and doesn't appreciably affect the precision of the
radial-velocity measurements, but calibration of the CCF width in term of
rotational velocity is substantially more difficult for the M4 template.

\begin{table}
\tabcolsep 0.1cm
\caption{Width and depth of the CCF of 6 stars with spectral type in the
 [M0V,M5.5V] range, for the M4 and K0 templates. 
 Spectral types are from Reid et al. (1995).}
\begin{tabular}{|l|ll|ll|} \hline
Spect. Type & ${\sigma}_{\rm M4}$(${\rm km\,s^{-1}}$) & $D_{\rm M4}$  &
${\sigma}_{\rm K0}$(${\rm km\,s^{-1}}$) & $D_{\rm K0}$ \\ \hline \hline 
M0 (Gl~424) & 4.71 & .094 & 5.00 & .113 \\
M1 (Gl~15A) & 4.59 & .104 & 4.99 & .106 \\
M2 (Gl~411) & 4.39 & .119 & 4.94 & .096 \\
M3 (Gl~725A) & 4.32 & .140 & 4.87 & .082 \\
M4 (Gl~699) & 4.21 & .159 & 4.96 & .062 \\
M5 (GJ~1002) & 4.26 & .176 & & \\ \hline
\end{tabular}
\label{caract}
\end{table}

For the slowly rotating stars in our sample, the above formula
predicts radial-velocity accuracies that range from 2~${\rm
m\,s^{-1}}$ for bright M0V (at typical S/N=200) to 70~${\rm
m\,s^{-1}}$ for faint M6V (at typical S/N=3). It is however clear that
the lowest values are somewhat optimistic, as the (excellent) long
term stability of ELODIE radial velocities is $\sim12~{\rm m\,s^{-1}}$
(Baranne et al. 1996).  We have therefore quadratically added ${\rm
12~m\,s^{-1}}$ to the internal errors to account for instrumental
stability.  A fiber optics light scrambler was recently added to
ELODIE to further stabilize the illumination of the spectrograph, and
has improved the long term radial-velocity stability to below $10~{\rm
m\,s^{-1}}$. We have however chosen not to use this device, as its
20\% light loss would degrade our overall measurement precision for
stars fainter than V~=~12.

Under some circumstances, the correction of the radial velocities to
the solar system barycentre represents an additional limitation on the
accuracy. At the latitude of Observatoire de Haute-Provence this
correction can vary by up to $1.5~{\rm m\,s^{-1}}$ per minute.  It
therefore significantly changes during the long exposure times (up to
1~hour) used for this programme, and it is important that the
effective time of the observation be accurately determined. For bright
stars a photomultiplicator monitors the fiber illumination of the
spectrograph to compute its exact value.  This P.M. is unfortunately
too noisy to be used for our programme stars, and we have instead to
use the mid-point of the observation.  Under good atmospheric
conditions this is not a problem, but when they are unstable (variable
seeing or passing clouds) the two values can differ by up to a few
minutes. In the worst case of unstable atmospheric conditions, a
1~hour exposure, and maximum variation of the barycentric correction,
this additional error can reach $15~{\rm m\,s^{-1}}$. It is usually
much smaller.

Similar, but much larger, errors can affect measurements of very short
period spectroscopic binaries (e.g. GJ~2069A, discussed below). Their
velocities can vary by as much as $300~{\rm m\,s^{-1}.min^{-1}}$ (for
P~=~1~day), and even small errors on the effective time of the
observation translate then into significant equivalent radial-velocity
errors.

Magnetic activity is another important potential limitation for
high-precision Doppler-velocity measurements of M~dwarfs. Many of them
have strong chromospheric activity, visible as H$_{\alpha}$ and Ca$^+$
H and K emission lines. The surface temperature of active stars is
inhomogeneous, and since hotter parts are brighter, the velocity
measured from the disk-integrated spectrum is biased towards the
velocity of any hot spot, or away from the velocity of any cold spot.
This effect is exacerbated when using a mismatched correlation
template, such as the K0 template for M dwarfs: the K0 template is a
better match to the hotter parts of the stellar surface and it
therefore further increases their relative weight in the measured
velocity.

On Julian day 2449978.5 (18.09.1995), a strong chromospheric flare in
Gl~873 provides an extreme example of activity-induced radial-velocity
errors, and illustrates the much better immunity of the M4 correlation
template. This generally active star presented on this date an
activity well above its usual level. The H$_{\alpha}$ emission line
showed a strong broad pedestal, characteristic (e.g. Jones et
al. 1996) of active stars observed within about an hour of a major
flare, and usually interpreted (Eason et al. 1992) as arising from
mass motion of gas clumps with a large volume filling factor. With the
K0 template, the radial velocity of Gl~873 for this date has a
precision of 150~${\rm m\,s^{-1}}$ but differs from the median Gl~873
value by 1.7~${\rm km\,s^{-1}}$. With the M4 template on the other
hand, the radial velocity precision for the same spectrum is improved
to 15~${\rm m\,s^{-1}}$, and the measured value is within only
40~${\rm m\,s^{-1}}$ of the median velocity. Given the extreme
character of the GL~873 event, 40~${\rm m\,s^{-1}}$ can thus be taken
as an upper bound on the maximum radial-velocity error that can result
from unusually high magnetic activity, when using the M4 mask.

% 47 exactement. En fait nettement davantage maintenant, mais pas revu depuis
% un an...
The examination of the $\sim$50 apparently single stars which have at
least 3 good measurements (internal precision better than
20~${\rm m\,s^{-1}}$) shows that about half of them have apparent excess
velocity dispersion at the $\sim{\rm 30~m\,s^{-1}}$ level. Gl~876 has been one
of those until we very recently accumulated enough measurements to
ascribe its velocity variability to a planetary companion (Delfosse et
al. 1998d; also Marcy et al. 1998). Most of these stars however have few
measurements, and it is thus not yet possible to determine how much of
the apparent dispersion is due to intrinsic or instrumental radial-velocity 
noise, and how much is due to yet unrecognized companions,
either planetary ones in short-period orbits or more massive distant
ones. The exact accuracy limits of this programme are thus still 
undetermined, but are better than $\sim30~{\rm m\,s^{-1}}$.

Double-lined spectroscopic binaries are in addition affected by a
phase-dependent systematic error source. The Gaussian profile
adjustment which is used to measure the radial velocity only
imperfectly describes the correlation profile of an M dwarf, whose
baseline shows some low level ($\sim$1\%) systematic undulations. For
a single star, or a single-lined binary, this has no effect on the
measured velocity, or perhaps introduces a small constant
zero-point offset. For double-lined binaries on the other hand, the
correlation peak of one star is superposed onto a systematically
different part of the other star's baseline, producing phase-dependent
velocity errors. If the velocity amplitudes are of the order of the
width of the the two correlation profiles, these are systematic errors
and they will bias the derived orbital elements, and in particular the 
amplitudes. When the amplitudes are sufficiently larger than the width of 
the profiles on the other hand, these velocity errors become random, and
the orbital elements then only have increased uncertainties but are unbiased.
% for stars measured with a high signal-to-noise ratio and
% TF: Pas necessaire en fait: a bas S/N le profil sera bruite et la 
% correction aussi, mais la precision de correction necessaire est 
% alors basse. La seule condition est reellement l'amplitude, plus
% le nombre de spectres et leur couverture de l'excursion en vitesse.
%
Under the same circumstances, and with enough measurements, it is also 
possible to separately determine the 
correlation profiles of the two components, and to correct the velocities
for the effect of their baseline wiggles, for instance with the iterative 
method used by Forveille et al. (1998) for Gl~570B. For smaller  
velocity amplitudes this method cannot be used, and one would have instead
to rely on the correlation profile of a matched spectral template, 
introducing some additional uncertainties. We haven't yet applied any
such correction except to Gl~570B, and SB2 velocities thus have errors of 
$\sim100~\rm{m\,s^{-1}}$ or larger (for the fainter component in particular).

\subsection{CORAVEL radial-velocity data}
65 bright (V{\less}12) stars in the sample had previously been
measured in the course of various observing programmes using the older
CORAVEL spectrographs (Baranne et al. 1979) on the 1-m Swiss telescope
(at Observatoire de Haute-Provence, France) and the 1.54-m Danish
telescope (La Silla Observatory, ESO, Chile).  The number of
measurements varies considerably, from 2-3 for stars examined in the
course of the HIPPARCOS follow-up survey (Udry et al. 1997), to over
100 for a few well studied previously known binaries. They have lower
precision than the ELODIE measurements (at best 300~${\rm m\,s^{-1}}$,
and often 1-2~${\rm km\,s^{-1}}$ for the present faint red stars),
but, when available, they extend the time base to typically
10~years. These data are thus systematically extracted from the
CORAVEL database for all programme stars.
% TF: Avons nous systematiquement verifie si la vitesse varie?
There is a significant radial velocity offset between CORAVEL
and ELODIE measurements (Udry et al. 1998), which must be corrected 
before they can be combined. For G-K dwarfs the offset depends upon 
both spectral type and radial velocity. Over the [M0V,~M3.5]
range  where we have common measurements the offset is independent of 
spectral type, but has a significant uncalibrated star to star
scatter at the $\sim600\,{\rm m\,s^{-1}}$ level. For well observed binary
stars, the CORAVEL offset is thus determined as part of the orbital
element adjustment.

\subsection{Adaptive optics data}

\subsubsection{Instrumental setup}
The observations were carried out at the 3.6-meter
Canada-France-Hawaii Telescope (CFHT) during several observing runs
from September 1996 to March 1998, using the CFHT Adaptive Optics
Bonnette (AOB) and two different infrared cameras. The AOB, also
called PUE'O, is a general purpose adaptive optics (AO) system. It is
mounted at the F/8 Cassegrain focus, and cameras and other
instruments are then attached to it (Arsenault et al. 1994, Rigaut et
al. 1998). It analyses the atmospheric turbulence with a 19 element
wavefront curvature sensor and corrects for it with a 19 degree of
freedom bimorph mirror. Modal control and continuous mode gain
optimization (Gendron \& L\'ena 1994; Rigaut et al. 1994) maximize the
correction quality for current atmospheric turbulence and guide star
magnitude. For our observations a dichroic mirror diverted the
visible light to the wavefront sensor while a science detector
recorded near-infrared light. The data were recorded with either MONICA, 
the Universit\'e de Montr\'eal Infrared Camera (Nadeau et al. 1994), or 
KIR, a new CFHT infrared camera developed to take full advantage of the 
AO corrected images produced by PUE'O (Doyon et al. 1998). 

MONICA was used for the commissioning of the AOB during the first
semester 1996 and for all science runs until November 1997. It is a
facility instrument based on a NICMOS3 256 $\times$ 256 detector, and was 
originally designed by the Universit\'e de Montr\'eal for the
Observatoire du Mont M\'egantic and CFHT F/8 Cassegrain focii. The camera 
was refitted with new optics for use at the F/20 output focus of
AOB. It produces a plate scale of 0\farcs034 per pixel, properly sampling
diffraction-limited images down to the J band (1.25 $\mu$m). The resulting
field size is $8.7\arcsec \times 8.7\arcsec$.

Since December 1997, MONICA has been replaced on PUE'O by KIR, a
new imaging camera which records a 16 time larger field on an HAWAII
1024 $\times$ 1024 HgCdTe array.  KIR has improved optical quality and
detector read-out noise, and therefore a significantly better
detectivity. The KIR plate scale is 0\farcs035 per pixel, for a total
field size of $36\arcsec~\times~36\arcsec$.

\subsubsection{Observations}
Sources were first examined for binarity with one filter, usually
H (1.65 $\mu$m). Under good to moderate seeing conditions H represents 
the best compromise between sensitivity, corrected image quality, and sky
brightness. Under worse seeing conditions the K filter was used
instead to maintain acceptable image quality. Sources which saturate
the detectors in the minimum available integration time through the 
H (1.65 $\mu$m) or K(2.23 $\mu$m) broad-band filter (brighter than K~=~7
under typical conditions) were observed through corresponding narrow-band 
filters, respectively [Fe$^+$] (1.65 $\mu$m) and H$_2$ (2.12 $\mu$m).
Whenever a target appeared double, it was further observed with
additional filters to determine relative colour indices. Integration
times per frame typically range between a few tenths of a second and a
few seconds. In order to improve the signal-to-noise ratio and to average
% the residual effects of atmospheric turbulence on the higher order modes, 
the residual uncorrected atmospheric turbulence, series of
$\sim$4~minute total integration times were accumulated in a four
position mosaic pattern. This observing sequence also allows to correctly
determine the sky background and to correct for detector cosmetic
defects.  Wavefront sensing was performed on the sources themselves,
which are always bright enough (R~$<$~14) to ensure diffraction-limited
images in H and K bands under standard Mauna Kea atmospheric
conditions (i.e.  seeing up to 1\arcsec). For most targets, PSF
calibration stars were observed under the same conditions to provide
input to parameter fitting (Section \ref{data}) and image
deconvolution. The atmospheric turbulence and AO correction for a
given set of observations were further characterized by simultaneously
recording the wavefront sensor measurements and deformable mirror
commands. An accurate synthetic PSF can be generated a posteriori from
these ancillary data, as described by V\'eran et
al. (1997). Astrometric calibration fields such as the central region of
the Trapezium Cluster in the Orion Nebula (McCaughrean and Stauffer 1994),
were observed to accurately determine the actual detector plate scale
and position angle (P.A.) origin. Flat-fields were obtained on the
dome and the sky for each filter.
 
\subsubsection{Data reduction}
\label{data}
\begin{figure*}
\psfig{height=12cm,file=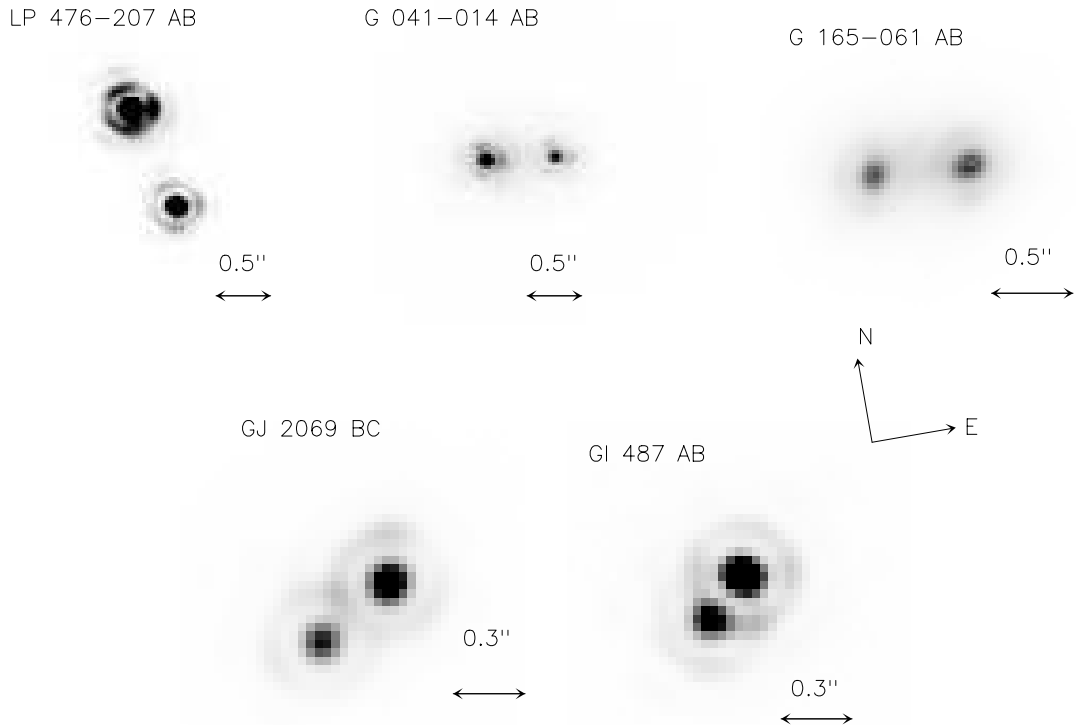,angle=-90}
\caption{Adaptive optics images for 5 of the new binaries.}
\label{oa}
\end{figure*}

% The data were reduced following standard image processing techniques
% for infrared observations, as described hereafter. 
For each filter, the raw images were median combined to produce sky
frames which were then subtracted from the raw data. Subsequent
reduction steps included flat-fielding, correction from the bad
pixels, and finally shift-and-add combinations of the corrected frames
into a final image. For resolved binary systems, the
separation, position angle and magnitude difference between the two
stars were determined using {\it uv} plane model fitting in the GILDAS
(Grenoble Image and Line Data Analysis System) software. With
approximate initial values of the positions of the two components
along with a PSF reference image, the fitting procedures gave as output
the flux and pixel coordinates of the primary and
secondary. Application of the astrometric calibrations then yields the
desired parameters. Images of the newly resolved binaries are
presented in Fig. \ref{oa}.

\section{New companions}

In this section we comment the new companions individually. A general view 
of their measurements (visual and spectroscopic) and status (new discovery, 
first orbit determination, etc.) is summarized in Table \ref{summarize}. 
If determined the orbital elements are given in Table \ref{tab_vr} and
some interesting visual parameters in Table \ref{tab_oa}.

\begin{table}
%XD APRES RETOUR DU REFEREE: preciser pour Gl876 que c'est pas
%une low mass star.
\caption{New low mass companions in the solar neighbourhood (the
  companion of Gl~876 is a planet).  Spectral types are from Reid et
  al. (1995), except for GJ~2130B which is taken from the CNS3
  catalogue. Parallaxes (in milliarcseconds) are taken from (a) the
  HIPPARCOS catalogue, (b) the Yale catalogue (Van Altena et al. 1995)
  or (c) are photometric parallaxes from the CNS3.}
\begin{tabular}{|l|ll|} \hline
Name & parallax (mas) & joint \\ 
 & & spectral type \\ \hline \hline
LP~476--207 AabB & 91.20$\pm$8.56$^{(a)}$  & M 4  \\
Gl~268.3~AB       & 81.05$\pm$2.42$^{(a)}$  & M 2.5  \\
GJ~2069~Aab      & 78.05$\pm$5.69$^{(a)}$  & M 3.5  \\
GJ~2069~BC        & 78.05$\pm$5.69$^{(a)}$  & M 4   \\
LHS~6158~AabB    & 224.0$\pm$36.0$^{(c)}$  & M 3.5  \\
Gl~381~AB         & 81.23$\pm$2.37$^{(a)}$  & M 2.5  \\
Gl~487~AabB       & 98.14$\pm$1.67$^{(a)}$  & M 3  \\
LHS~2887~AB       & 62.2$\pm$13.1$^{(b)}$   & M 4  \\
G~203--047~ab      &  137.84$\pm$8.95$^{(a)}$   & M 3.5  \\
GJ~2130~Bab      & 161.77$\pm$11.29$^{(a)}$& M 2.5 \\ 
Gl~829~ab         & 148.29$\pm$1.85$^{(a)}$ & M 3.5  \\
Gl~876~ab        & 212.69$\pm$2.10$^{(a)}$ & M 4   \\
Gl~896~Aab       & 160.06$\pm$2.81$^{(a)}$ & M 3.5  \\ 
Gl~896~Bab       & 160.06$\pm$2.81$^{(a)}$ & M 4.5  \\ \hline
\end{tabular}
\end{table}

\begin{figure*}
\begin{tabular}{cc}
\psfig{height=5.5cm,file=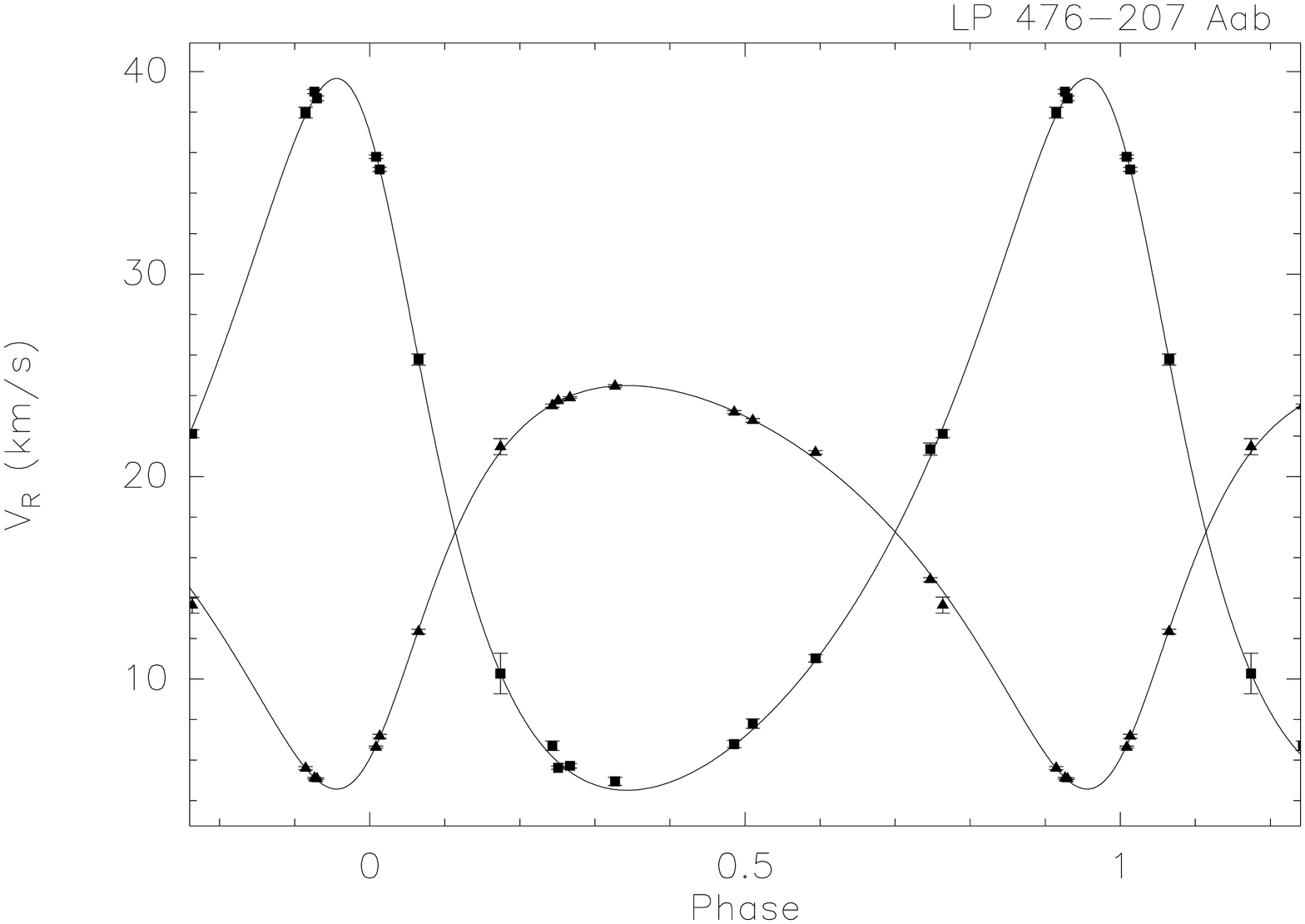,angle=-90} &
\psfig{height=5.5cm,file=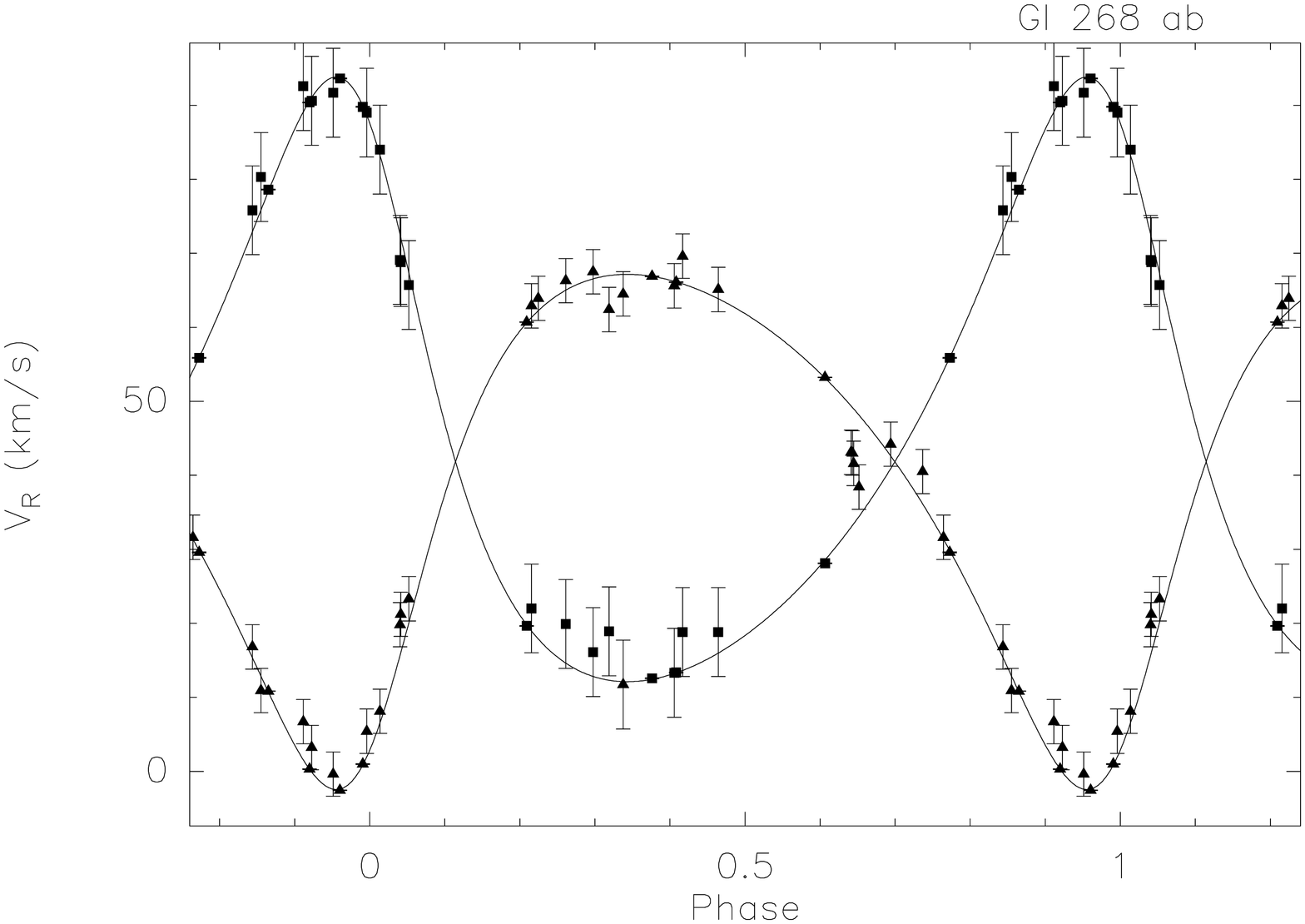,angle=-90} \\
\psfig{height=5.5cm,file=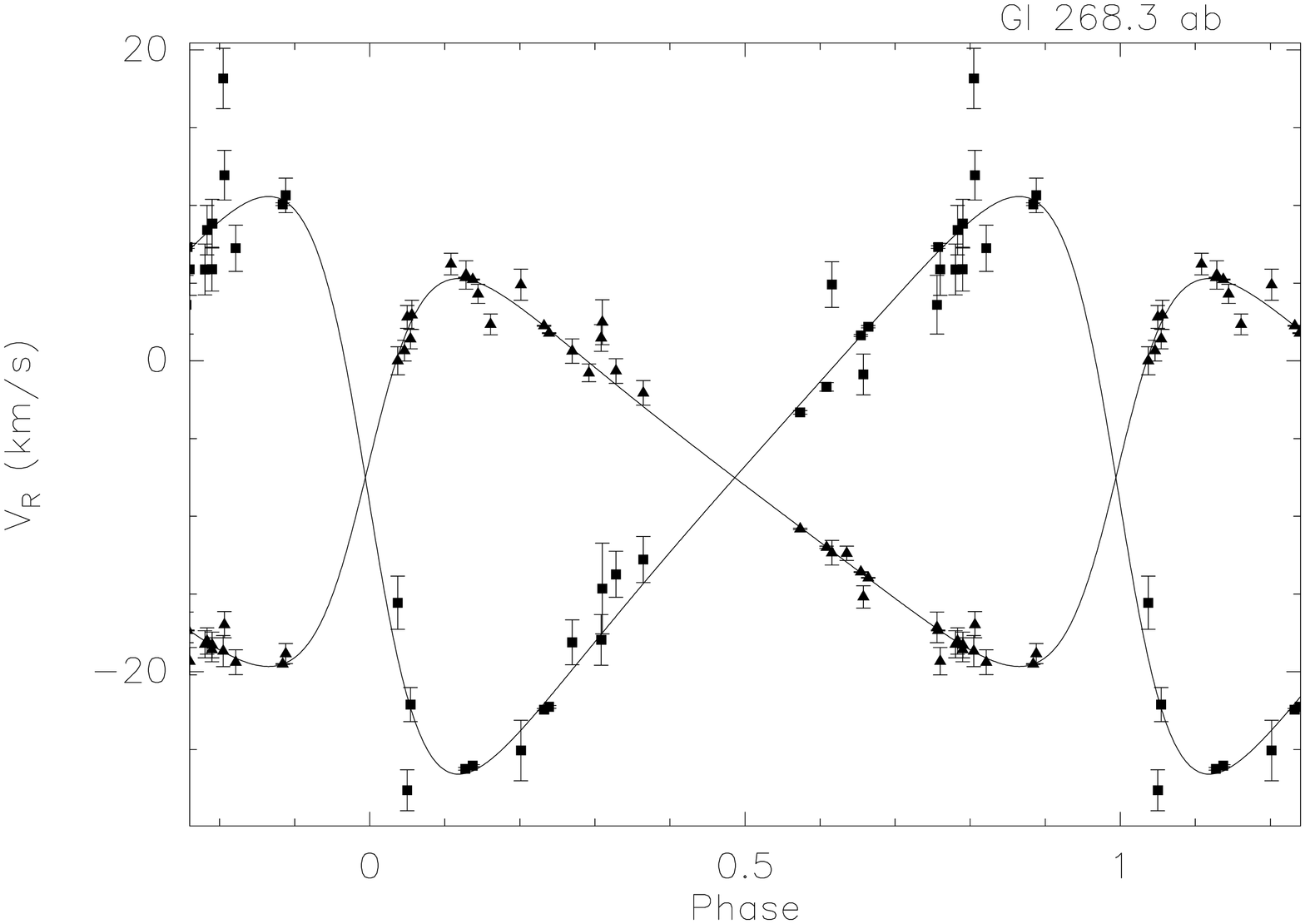,angle=-90} &
\psfig{height=5.5cm,file=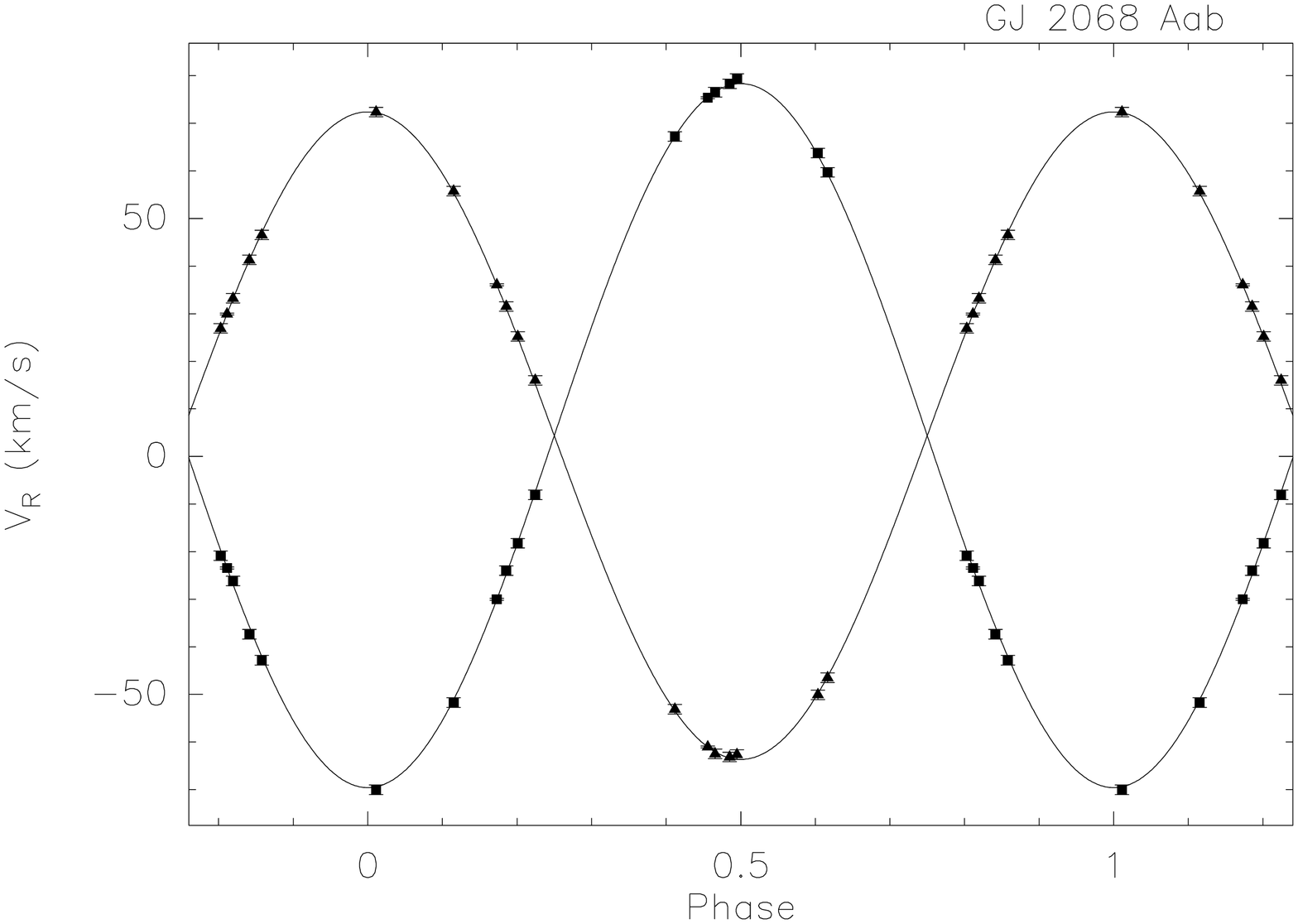,angle=-90} \\
\psfig{height=5.5cm,file=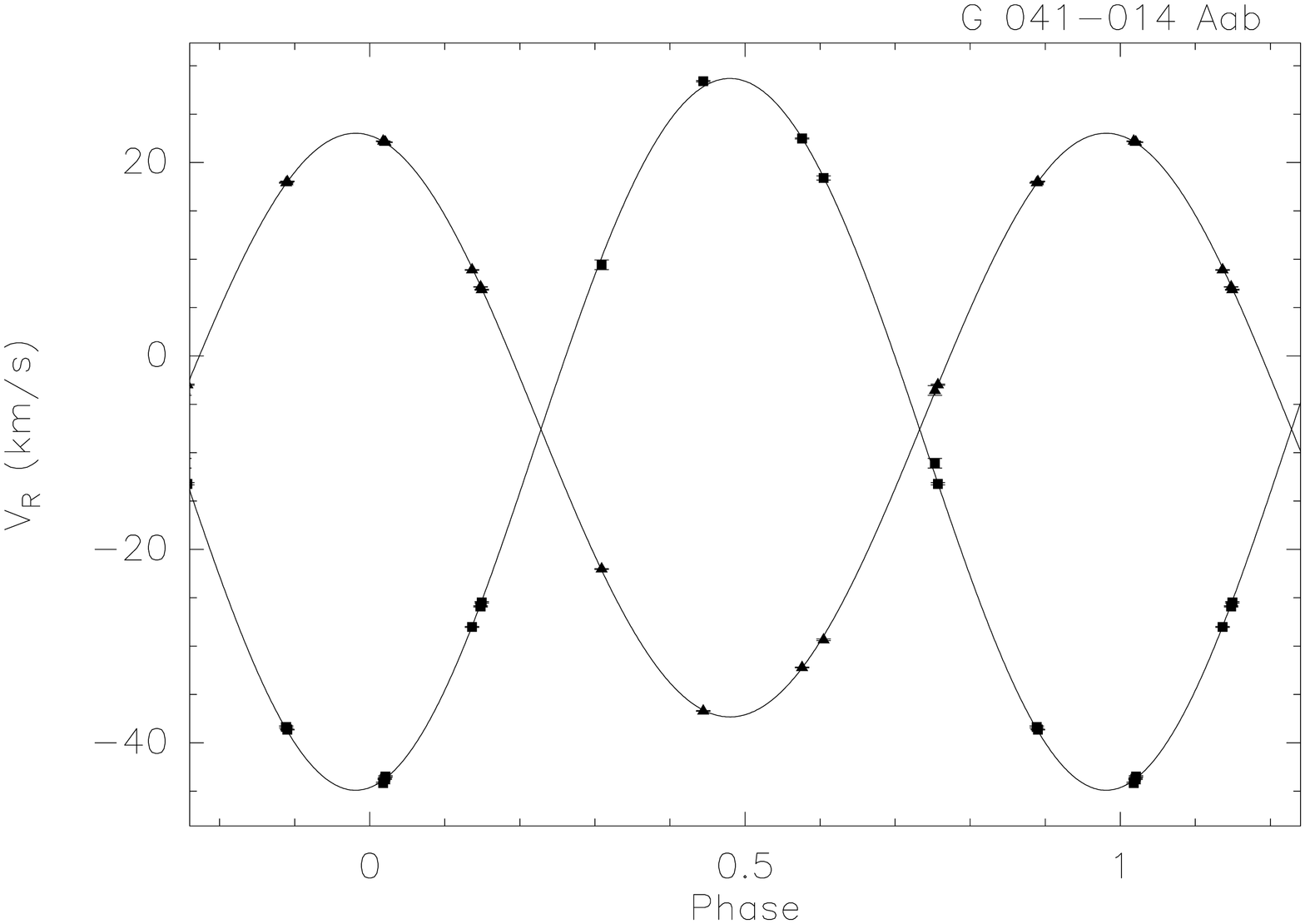,angle=-90} &
\psfig{height=5.5cm,file=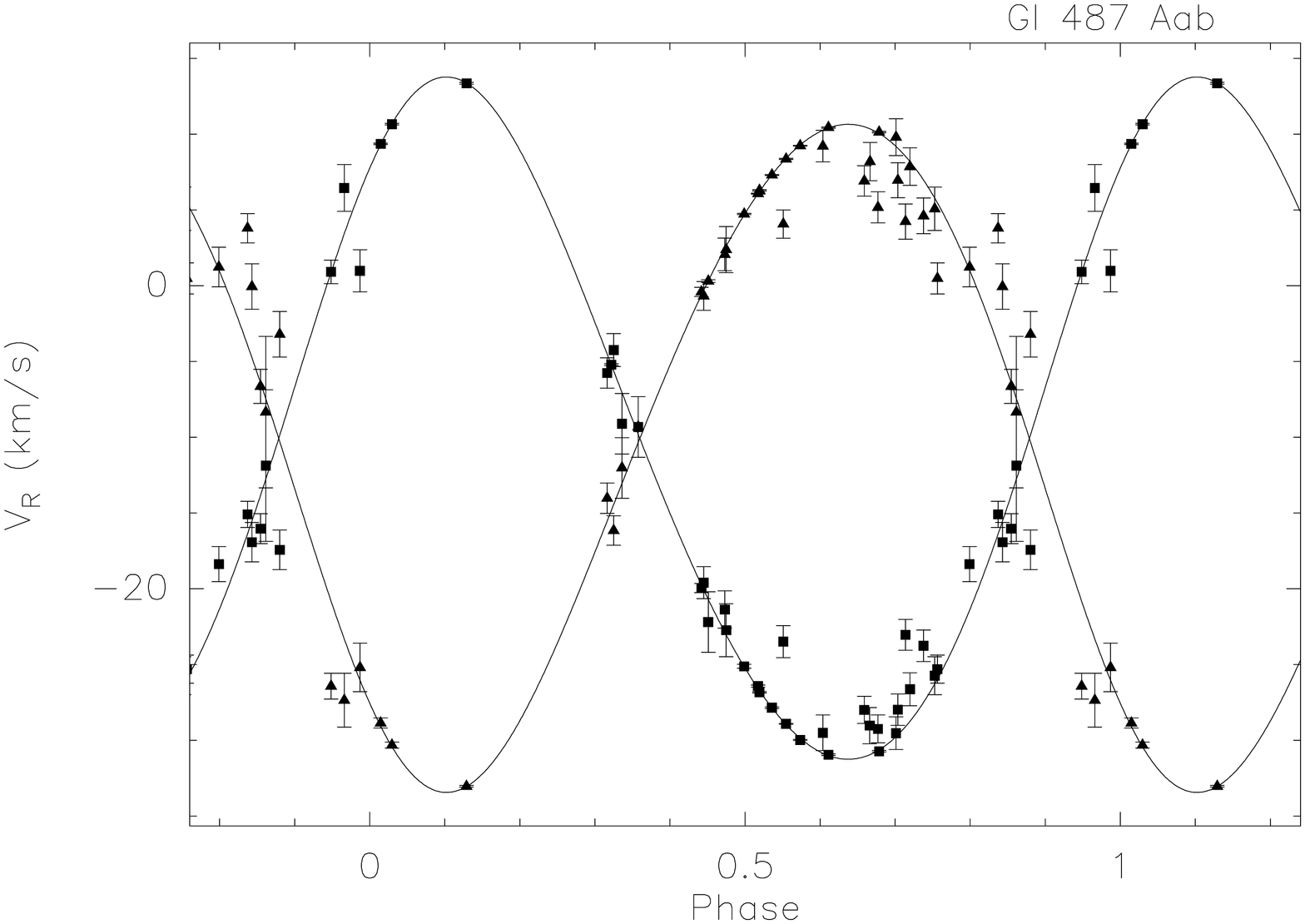,angle=-90} \\
\psfig{height=5.5cm,file=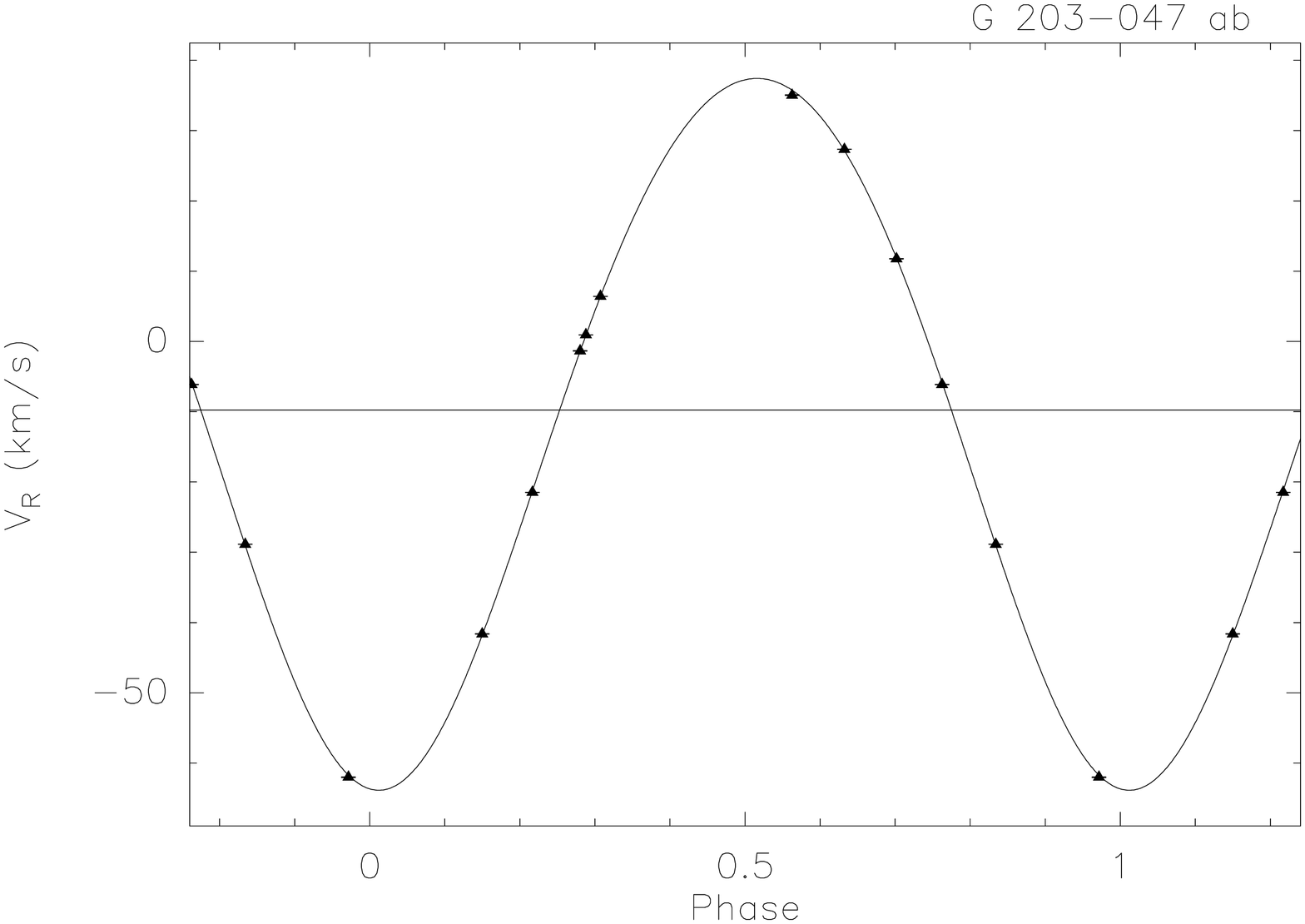,angle=-90} &
\psfig{height=5.5cm,file=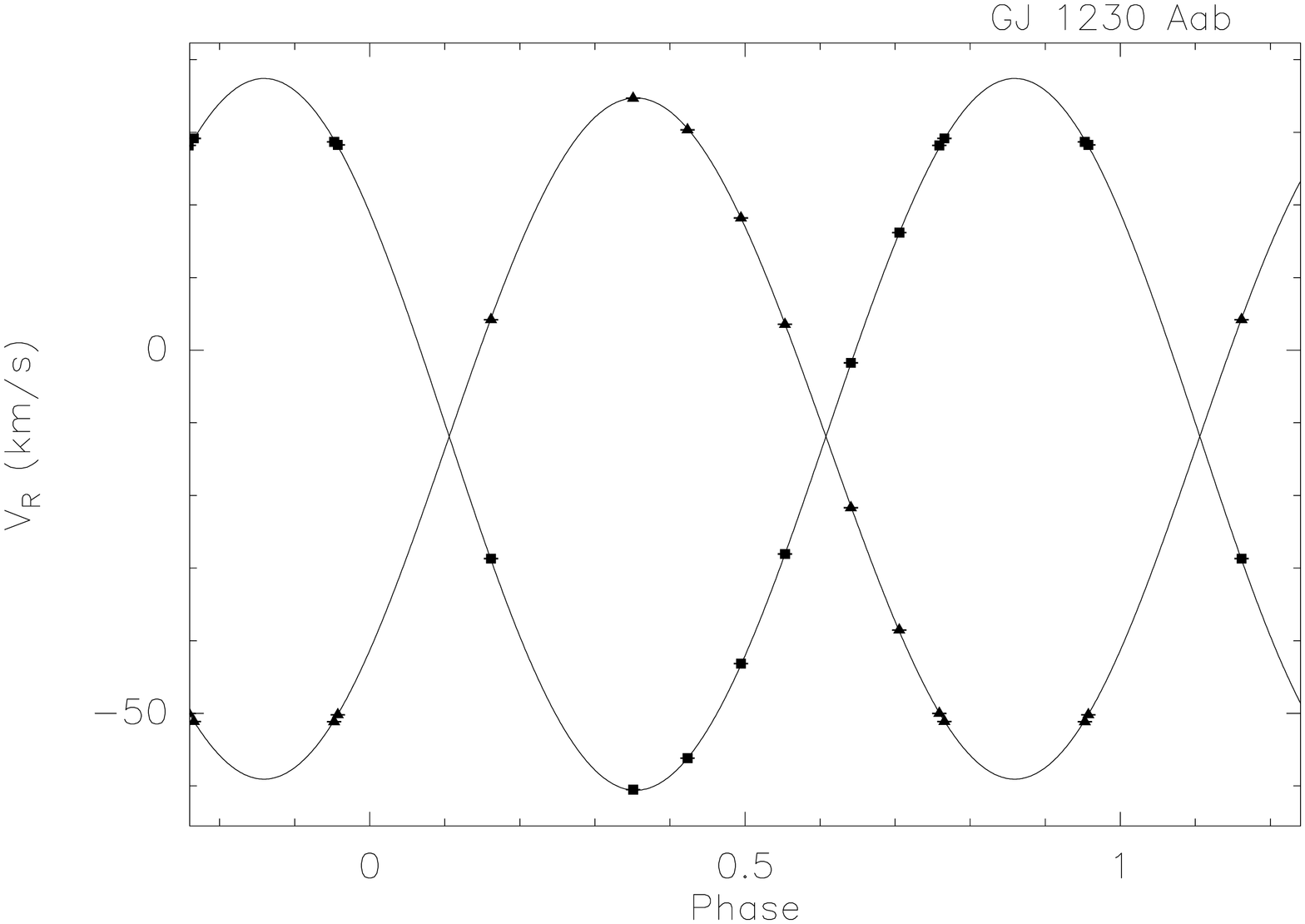,angle=-90} \\
\end{tabular}
\caption{Radial-velocity orbits. Triangles correspond to radial velocities
  of the primary star and squares to those of the secondary star.}
\end{figure*}

\begin{figure*}
\begin{tabular}{cc}
\psfig{height=5.5cm,file=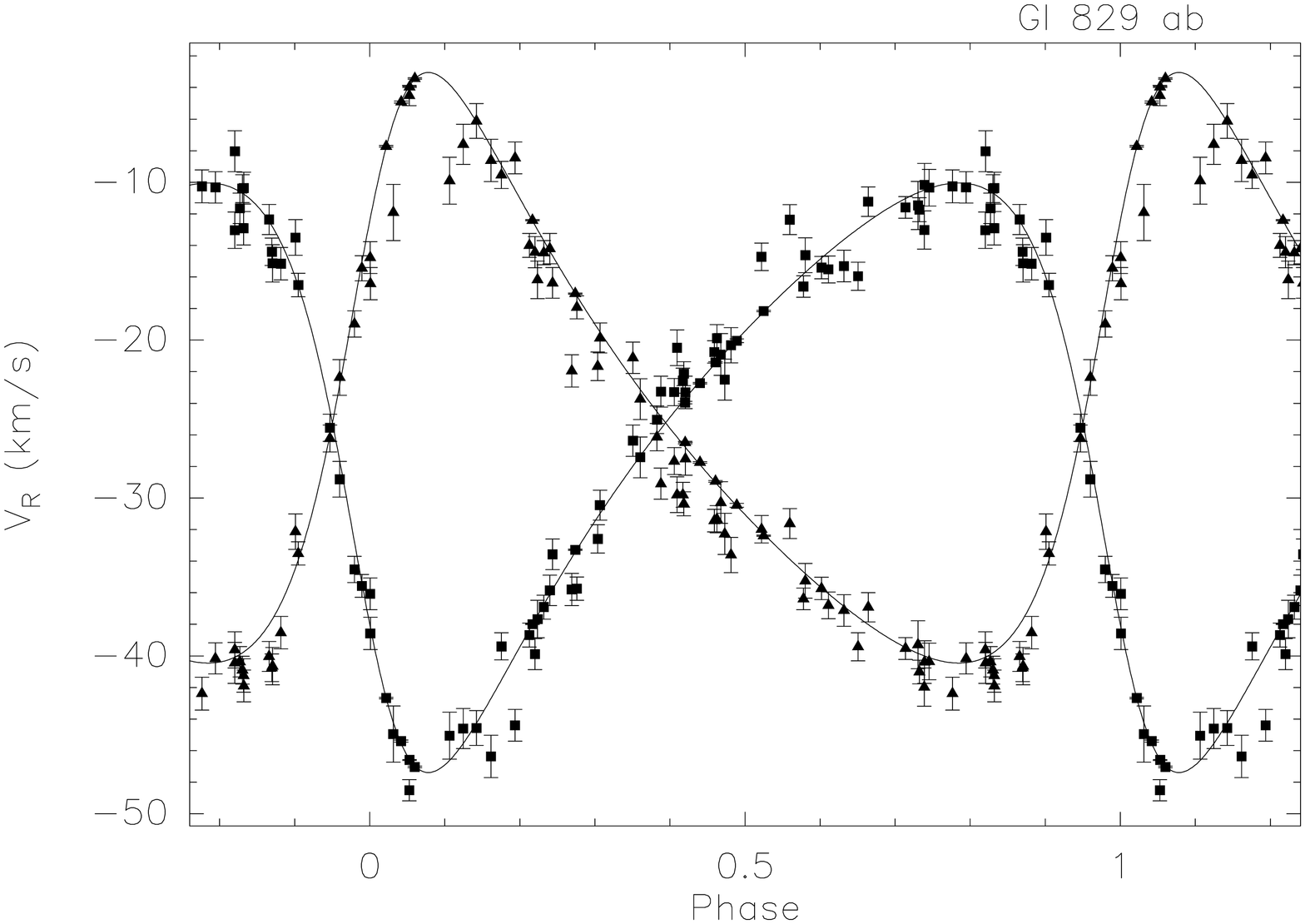,angle=-90} &
\psfig{height=5.5cm,file=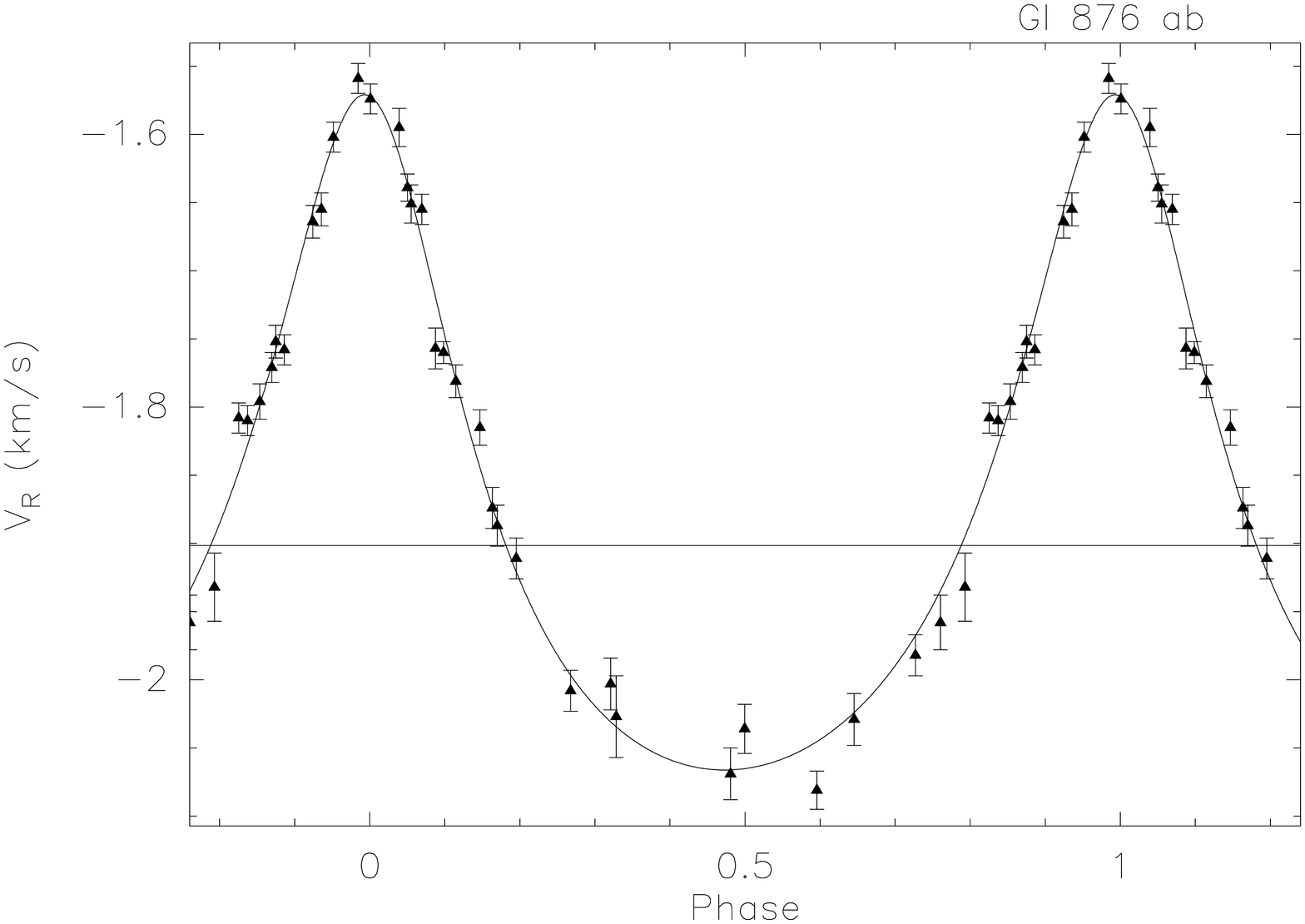,angle=-90} \\
%\hspace{0.1cm}
%  \begin{minipage}[b]{7cm}
%  \caption{Radial velocity orbits. The triangle indicate primary and the
%square
%  indicate secondary.
%    \vspace{1cm}} 
%  \end{minipage}
\end{tabular}
\centerline{{\bf Fig. 3.} (continued) 
%Radial velocity orbits. Triangles correspond to radial velocities
%of the primary star and squares to those of the secondary star.
}
\end{figure*}

\subsection{LP~476--207~AabB}

LP 476--207 is a new triple system. 
Adaptive optics images show a 0.97'' separation pair with a K-band 
magnitude difference of 0.9 (Table \ref{tab_oa}), 
confirming a recent speckle detection of 
this outer component by Henry et al. (1997). The brightest
component of the visual pair, LP~476--207~A, is a new large amplitude 
double-lined spectroscopic binary, with a period of $P{\sim}12~$days. 
The visual pair was also detected by HIPPARCOS (ESA, 1997), which lists a 
separation of 0.68''. There has thus been significant orbital motion over 
the last $\sim$5~years.  
%{\em  Distance \\ }
% {\em Also detected by HIPPARCOS,\\ 
% Ncomp MultFlag Source Qual m  theta    rho    e_rho   dHp  e_dHp 
%    2 C               D    AB   178    0.677  0.022   1.00  0.20

\subsection{Gl~268~ab}
Gl~268 is a previously known double-lined binary on a 10.43-day orbit
(Tomkin and Pettersen 1986). Our new spectroscopic measurements
provide substantially improved orbital elements.

\subsection{Gl~268.3~AB}
Gl~268.3 is a new double-lined spectroscopic binary. With respectively 
29 and 10 CORAVEL and ELODIE measurements, spanning over 15 years, the
radial-velocity orbit is extremely well determined. The
period is 304.35$\pm$0.25 days. K-band adaptive optics images partly 
resolve this system into a $\sim$0.1'' binary with a small magnitude
difference (Table \ref{tab_oa}).
% pas de diff de mag pour l'instant 
%with $\sim$X.Y magnitude
%difference at K band. 
Within one or two years Gl~268.3 will thus provide two precise mass 
determinations at a spectral type of M2.5V.
% Pas detectee comme binaire par HIPPARCOS: configuration serree et faible
% difference de magnitude.

\subsection{GJ~2069~AabBC}

%(voir les precisions sur vr de GJ~2069AB, derive et correction..)

Previously known as a wide ($\sim12''$) visual binary, the GJ~2069
system is actually quadruple.  Adaptive optics images resolve GJ~2069B
into a 0.36'' binary with a K-band magnitude difference of 0.45 (Table
\ref{tab_oa}). 
The 5~radial-velocity measurements obtained over
$\sim$850~days show a linear drift of $\sim600~\rm{m\,s^{-1}}$. 
% En fait on a une mesure, 700jours de trou, puis 4 mesures en 120jours.
% Est-ce suffisant pour parler d'une derive lineaire???
The period is thus probably long and the small visual separation may be
due to projection.

GJ~2069A is a new short-period (P~=~2.8~days) detached double-lined 
eclipsing binary, only the third one known with substantially sub-solar 
masses. 
%(with YY~Gem and CM~Dra).
Delfosse et al (1998c) discuss in detail the very accurate 
radial-velocity orbit and a preliminary light curve.  They derive
individual masses (0.430 and 0.396 {\Msol}) with 0.4\%
accuracy. Both components are clearly sub-luminous for their masses, and
their spectral types are later than expected, compared with both solar
metallicity stellar models and previous stellar mass
determinations. This indicates a metal-rich composition.

\subsection{LHS~6158 (G~041--014)~AabB}
Recently shown to be a short-period double-line spectroscopic binary
(Reid \& Gizis 1997), G~41-14 is actually a triple system. We have
determined the orbital elements ($P=7.6$~days) of the close
spectroscopic pair, and discovered a third component in AO images,
at a separation of $0.62''$ and with a K-band magnitude difference of
0.5 (Table \ref{tab_oa}). The probable period is $\sim$10 years and 
a mass determination is thus a mid-term prospect only.

\subsection{Gl~381~AB}
This new long-period double-line spectroscopic binary is also well
resolved into a 0.18'' pair ($\Delta$K=0.95~mag) in AO images (Table 
\ref{tab_oa}). The period is relatively well determined at ${\sim}2845$~days, 
but the CORAVEL measurements do not separate the two components well and the
ELODIE data only cover a small fraction of the orbit. All other orbital 
elements are thus still uncertain.  ELODIE will easily separate the two 
stars at periastron. This spectroscopic~+~visual pair will then
provide accurate mass measurements, within about 5~years. 
%
%For this binary, the discussion about the orbital parameters 
%is deferred to a forthcoming paper (Perrier et al., in preparation). 
% ref a futur papier?
% Pas vue par HIPPARCOS

\subsection{Gl~487~AabB}
Gl~487 is a new triple-line spectroscopic system. The short period
(for Gl~487Aab) is 54.07 days and the long period is approximately
3000~days. The outer system (Gl~487~AB) is also resolved in
AO images with a separation of $0.23''$ and $\Delta$K=0.7~mag 
(Table \ref{tab_oa}). This system has the potential to provide three 
accurate mass
determinations, but its complete analysis is deferred to a forthcoming
paper (Delfosse et al. in preparation) dealing with triple systems.
% Pas vue double par HIPPARCOS

\subsection{LHS~2887 (G165--061)~AB}

The probable period of this new double-line spectroscopic binary is a
few years (1500~days for a very preliminary orbit). It is also well 
resolved in AO images, with a separation
of 0.58'' (Table \ref{tab_oa}) and will eventually provide accurate masses.
%{\em Distance} 
%{\em Verifier si les ordres de grandeurs collent entre periode et 
%separation, ou si c'est plus probablement une triple}
% Pas dans HIPPARCOS

\subsection{Gl~644}
Our spectroscopic measurements confirm that the speckle binary Gl~644
(Tokovinin et Ismailov
1988) is a triple system, as previously shown by Eggen (1978) and
Pettersen et al. (1984). We have decoupled the two orbits,
and obtained preliminary values for
the masses of the three stars. Their accuracy would
however benefit from additional measurements and their discussion is 
thus deferred to a forthcoming paper (Delfosse et al., in preparation).  

\begin{table*}
\caption{Measurement summary: COR, ELO and PUE'O are the number of
measurements obtained with CORAVEL, ELODIE and PUE'O, respectively. Gl~644 and
Gl~866 are not included here, since their discussion is deferred to a 
forthcoming paper (Delfosse et al. in preparation). To date (September 1998),
this programme has identified 13 new stellar companions and 1 planet 
orbiting M dwarfs of the solar neighbourhood. Orbital elements have been 
determined for six of them, as well as for 3 previously known binaries
with undetermined orbits (LHS~6158~Aab and G~203--047~ab, Reid and Gizis
(1997); GJ~1230~Aab, Gizis and Reid (1996)).}
\begin{tabular}{|l|ccc|cccc|} 
\hline
%Name & N$_{\mbox{\rm\tiny{COR}}}$ & N$_{\mbox{\rm\tiny{ELO}}}$ 
Name            & \multicolumn{1}{c}{COR}   & \multicolumn{1}{c}{ELO}   & PUE'O
 & discovered by
& first orbit   & orbit         & In the 9-pc \\
                &       &       &       & this programme& determination &
improvement     & sample      \\ \hline \hline
LP~476-207~Aab  & 0     & 16    & 3     & yes   
& yes           & --            & no    \\
LP~476-207~AB   & 0     & 16    & 3     & no            & no
        & no            & no    \\
Gl~268~ab       & 0     & 9     & 1     & no            & no
        & yes           & yes   \\
Gl~268.3~AB     & 29    & 10    & 3     & yes           &
yes             & --            & no    \\
GJ~2069~Aab     & 0     & 18    & 1     & yes           &
yes             & --            & no    \\
GJ~2069~BC      & 0     & 5     & 1     & yes           & no
        & no            & no    \\
LHS~6158~Aab    & 0     & 14    & 2     & no            &
yes             & --            & yes   \\
LHS~6158~AB     & 0     & 14    & 2     & yes           & no        
   & no                 & yes   \\
Gl~381~AB       & 27    & 4     & 2     & yes           & no  
         & no           & no    \\
Gl~487~Aab      & 24    & 18    & 5     & yes           & yes  
        & --            & no    \\
Gl~487~AB       & 24    & 18    & 5     & yes           & no
        & no            & no    \\
LHS~2887~AB     & 0     & 5     & 2     & yes           & no
        & no            & no    \\
%Gl~644
G~203-047~ab    & 0     & 12    & 1     & no            &
yes             & --            & yes   \\
GJ~2130~Bab     & 0     & 1     & 0     & yes           & no
        & no            & yes   \\
GJ~1230~Aab     & 0     & 11    & 0     & no            &
yes             & --            & yes   \\
Gl~829 ab       & 67    & 11    & 1     & yes           &
yes             & --            & yes   \\
%Gl~866
Gl~876~ab       & 0     & 32    & 1     & yes           & yes   
& --            & yes   \\
Gl~896~Aab      & 0     & 9     & 1     & yes           & no 
        & --            & yes   \\
GL~896~Bab      & 0     & 6     & 1     & yes           & no
        & --            & yes   \\  \hline
\end{tabular}
\label{summarize}
\end{table*}

\begin{table*}
\caption{Orbital elements and M\,${\sin}i$ for the newly determined 
radial-velocity orbits. The inclination of GJ~2069 Aab is accurately known 
and we thus list the actual masses instead of M\,${\sin}i$.
G203--47 and Gl~876 are single-lined binaries (all others are double-lined), 
so we list $f_1(M)$ as well as estimated M$_{2}~\sin^3{i}$ for assumed
primary masses of, respectively, 0.35 and 0.30~{\Msol}.}
\begin{tabular}{|l|lllllll|ll|} 
\hline
%Name & N$_{\mbox{\rm\tiny{COR}}}$ & N$_{\mbox{\rm\tiny{ELO}}}$ 
Name            & \multicolumn{1}{c}{P}             & \multicolumn{1}{c}{T$_0$}

& \multicolumn{1}{c}{e}             &
\multicolumn{1}{c}{$\omega$}        & \multicolumn{1}{c}{K$_1$}         &
\multicolumn{1}{c}{K$_2$}         & \multicolumn{1}{c}{V$_0$}         &
\multicolumn{2}{c|}{mass determination} \\ 
                & \multicolumn{1}{c}{(days)}        &
\multicolumn{1}{c}{(Julian day)}  &
                               &
        &       \multicolumn{1}{c}{(${\rm km\,s^{-1}}$)}       
&\multicolumn{1}{c}{(${\rm km\,s^{-1}}$)}&
         \multicolumn{1}{c}{(${\rm km\,s^{-1}}$)}  &
\multicolumn{2}{c|}{(\Msol)}                      \\ \hline \hline
\multicolumn{8}{|l|}{Double-lined binaries: ${\rm M}_1~{\times}\sin~i$ and 
${\rm M}_2~{\times}\sin~i$}                             &
M$_{1}~\sin^3{i}$       &M$_{2}~\sin^3{i}$      \\ \hline
LP~476-207~Aab  & 11.9623       & 49799.47      & .323          &
212.0   & 9.96          & 17.57         & 17.26         & .014 
                & .008                  \\
                & $\pm$.0005    & $\pm$0.04     & $\pm$.006     &
$\pm$0.6        & $\pm$0.03     & $\pm$0.07     & $\pm$0.03     &
$\pm$.0001              & $\pm$0.0001           \\ \hline
Gl~268~ab       & 10.4265       & 50149.902     & .321          & 212.1 
& 34.81         & 40.86         & 41.83         & .215                  &
.183                    \\ 
                & $\pm$.00002   & $\pm$0.008    & $\pm$.001     &
$\pm$0.3        & $\pm$0.04     & $\pm$0.06     & $\pm$0.03     &
$\pm$.001               & $\pm$.001             \\ \hline
Gl~268.3~AB     & 304.35        & 48826.0       & .399          & 273.8 
& 12.47         & 18.6          & -7.51         & .435          
& .292                  \\
                & $\pm$.25      & $\pm$1.5      & $\pm$.008     &
$\pm$0.9        & $\pm$0.08     & $\pm$0.2      & $\pm$0.05     &
$\pm$.010               & $\pm$.006             \\ \hline
LHS~6158~Aab    & 7.5555        & 50471.2       & .014          & 7.0   
& 30.15         & 36.79         & -7.57         & .129                  &
.106                    \\
                & $\pm$.0002    & $\pm$0.2      & $\pm$.002     &
$\pm$9.3        & $\pm$0.05     & $\pm$0.09     & $\pm$0.04     &
$\pm$.001               & $\pm$.001             \\ \hline 
Gl~487~Aab      & 54.075        & 50506.2       & .081          & 137.0  
        & 22.0          & 22.5          &  -10.1        & .247          
& .242                  \\
                & $\pm$.006     & $\pm$0.5      & $\pm$.005     &
$\pm$3.0        & $\pm$0.5      & $\pm$0.5      & $\pm$0.4      &
$\pm$.014               & $\pm$.013             \\ \hline
GJ~1230~Aab     & 5.06880       & 50643.7       & .009          & 230.0
        & 46.9          & 49.0          & -11.88        & .237  
        & .226                  \\ 
                & $\pm$.00005   & $\pm$0.2      & $\pm$.001     &
$\pm$10.0       & $\pm$0.1      & $\pm$0.1      & $\pm$0.05     &
$\pm$.001               & $\pm$.001             \\ \hline
Gl~829~ab       & 53.221        & 48980.2       & .374          & 300.0
        & 18.7          & 18.7          & -25.23        & .114   
        & .114                  \\ 
                & $\pm$.004     & $\pm$0.2      & $\pm$.004     &
$\pm$1.0        & $\pm$0.1      & $\pm$0.1      & $\pm$0.06     &
$\pm$.001               & $\pm$.001             \\ \hline
\multicolumn{8}{|l|}{Single-lined binaries: $f_1(M)$ and an estimate of 
M$_{2}~{\sin}^3{i}$}                                    & $f_1(M)$             
&
M$_2~{\sin}^3{i}$       \\ \hline
G203-47~Aab     & 14.7136       & 50500.8       & .068          & 175.0
        & 50.6          & --            & -9.7          & 0.2 
                & $\sim$~0.5            \\
                & $\pm$.0005    & $\pm$0.1      & $\pm$.004     &
$\pm$3.0        & $\pm$0.2      & --            & $\pm$.2       & --
                & --                    \\ \hline
Gl~876~ab       & 61.1          & 50661.7       & .33   
& 5.0           & 0.247         & --            & -1.901
& 8.10$^8$              & $\sim$~0.002          \\ 
                & $\pm$.2       & $\pm$1.5      & $\pm$.02      & $\pm$5.0
& $\pm$0.006    & --            & $\pm$0.005    & --    
        & --                    \\ \hline
\multicolumn{8}{|l|}{Double-lined eclipsing binary: complete determination of
M$_1$ and M$_2$}                                        & M$_1$         
& M$_2$                 \\ \hline
GJ~2069~Aab     & 2.771472      & 50207.8128    & .0            & -- 
        & 68.03         & 73.06         & 4.36          & .430 
                & .396                  \\
                & $\pm$.000004  & $\pm$0.0009   & $\pm$.003 
& --            & $\pm$0.09     & $\pm$0.09     & $\pm$0.04
& $\pm$.001             & $\pm$.001             \\ \hline
\end{tabular}
\label{tab_vr}
\end{table*}

\subsection{G~203--047~ab}
This M3.5V star was previously noted as a single-line spectroscopic binary
by Reid \& Gizis (1997). We have obtained orbital elements, with a 15-day 
period, a large velocity semi-amplitude of 50~${\rm km\,s^{-1}}$, and a small
and only marginally significant eccentricity of 0.07. It is listed in
the HIPPARCOS catalogue as a probable short-period astrometric binary
without orbital solution. Now knowing the period, it would be of interest 
to reanalyse the HIPPARCOS data, and extract the inclination. 
The orbital elements result in a large mass function, 
% $\over{(M_2~\sin{i})^3}{(M_1+M_2)^2}~=~0.2{\Msol}$, 
${(M_2~\sin{i})^3}/{(M_1+M_2)^2}~=~0.2{\Msol}$, 
and therefore imply a quite massive secondary. Adopting the Baraffe et al 
(1998) theoretical mass-luminosity relation, the mass of the M3.5V primary 
(M$_{\rm I}$=8.87, Figueras et al. 1990)
is 0.30-0.35{\Msol},
% 0.2 [M2.5,3.5] 0.15 [M3.5,M4.5]
and the measured mass function thus requires that the secondary 
component be more massive than $\sim$0.5{\Msol}.    
% M1=0.15Msol  ==> M2min=0.38 (K1=20)
% M1=0.20Msol  ==> M2min=0.42 (K1=24)
% M1=0.30Msol  ==> M2min=0.5
% M1=0.35Msol  ==> M2min=0.55
This clearly excludes a single-main sequence star, which would have a
spectral type earlier than M1V and would dominate the observed light.
A (very) short-period main-sequence binary is similarly excluded,
because its brighter member would still need to be at least about as 
massive as the primary, and would be easily visible in the spectrum. The
companion to G~203-047 must thus be a degenerate star. It is most
likely a white dwarf, since visible photometry of G~203-47 (Figueras
et al. 1990) shows a $\sim$0.4~U-B excess over the colour of stars
with the same R-I colour or spectral type (Leggett 1992). UV
spectroscopy would easily ascertain the exact characteristics of the
white dwarf. The small semi-major axis of the present orbit
($a_1\sin{i}~=~14.5\Rsol$, or 0.05 AU) implies that G~203-47a must have 
been, at previous stages, in a contact configuration with the AGB
progenitor of its white dwarf companion. It is therefore a very nearby
member of the pre-cataclysmic variable family (e.g. Ritter \& Kolb 1998), 
but the timescales for its evolution into a CV is extremely long. 
It should have accreted
some nucleosynthesis products dredged up to the surface of its
previous AGB companion, and its composition may thus be peculiar.
% Is accreted mass large enough to significantly affect the composition in 
% spite of dilution in the fully mixed M3.5V star??
%
% G~203-47 thus
% represents an interesting observational constraint on the maximum 
% separation for final merging of common envelope systems.
% TF: Pas vraiment en fait, des cas nettement plus serres sont connus
\\

\subsection{GJ~2130~ABab}
% Dans HIPPARCOS 
At ${\delta}=-32^{\circ}$, the little studied GJ~2130 system is far below
the declination limit of the main sample and was observed as part of
a possible southern extension of the programme. GJ~2130B, the fainter 
component of this wide (25{\arcsec} separation) visual binary turns out to be 
a double-lined spectroscopic binary, and the system is thus triple. To date 
we don't have enough radial-velocity data to attempt an orbital solution.

\subsection{GJ~1230~AabB}
% Pas dans HIPPARCOS (juste une etoile F (double) a 3').
GJ~1230A, the brighter component of the wide (5{\arcsec} separation)
GJ~1230AB visual binary is a double-lined spectroscopic binary (Gizis
and Reid, 1996), for which we determine first orbital elements. The
period is 5.1 days.

\subsection{Gl~829~ab}
Gl~829 was mentioned as a possible double-lined spectroscopic binary
by Marcy et al. (1987). It is clearly seen as such by both
CORAVEL (67 measurements) and ELODIE (11 measurements) and the orbital
elements are well constrained. The orbital period is 53.2~days and the
mass ratio is very close to 1.
% Pas vue binaire par HIPPARCOS, serree et luminosites egales

\subsection{Gl~866}
We confirm that the astrometric and
speckle binary Gl~866 (Leinert et al. 1986) is a triple system, as
long suspected from the mass excess of one of the speckle
components (Leinert et al. 1990). It is seen as a triple-lined
spectroscopic system in the ELODIE spectra. The extensive speckle
coverage of its outer orbit (Leinert et al. 1990) makes it a prime
candidate for an accurate mass determination. As for Gl~644, the two 
decoupled orbits and the resulting masses will be discussed in a forthcoming
paper (Delfosse et al., in preparation).

% {\em Ou Leinert et al.  pour 866??}

% 644 resolue par HIPPARCOS, qui donne curieusement une separation unique.
% La parallaxe est assez differente de celle de 643 (20mas, pour des sigmas
% sur chaque de 4mas). Probleme du a l'orbite (P=625 jours)?? A dicuster
% avec un specialiste.
% 866 pas dans HIPPARCOS

% A garder dans cet article??
\subsection{Gl~876~ab}
High precision radial-velocity observations of the nearby M4 dwarf Gl~876
with the Observatoire de Haute-Provence 1.93-m telescope and the new
1.20-m Swiss telescope at La Silla indicate the presence of a Jovian
mass companion to this star. The orbital fit to the data gives a period
of 60.96~days, a velocity amplitude of 248~m\,s$^{-1}$ and an eccentricity
of 0.34. Assuming that Gl~876 has a mass of 0.3~{{\Msol}}, the mass function
implies a mass for the companion of 2/$\sin{i}$ Jupiter masses. Delfosse et al.
(1998d) and Marcy et al. (1998) both discuss this interesting system in detail.

\subsection{Gl~896~AabBab}
The Gl~896 system, well known as a wide binary, is actually quadruple:
both members of the visual pair are new single-lined spectroscopic
binaries. Their probable periods are of the order of a few years but still
undetermined. We
have observed both of them with adaptive optics in December 1997 and 
not resolved either. The luminosity contrast of the companions is thus 
probably large.
% HIPPARCOS ne voit que la paire AB

\begin{table}
\caption{Projected separation and luminosity contrast at K band, for 
binaries resolved by adaptive optic imaging.}
\begin{tabular}{|l|l|l|l|} \hline
Name & Separation & ${\Delta}$K & Observing \\ 
     &            & (mag)    & date \\ \hline \hline 
LP~476-207~AB & 0.97"$\pm$0.01" & 0.9$\pm$0.05  & Jan 1997 \\
Gl~268.3~AB   & \multicolumn{2}{l|}{marginally resolved ${\sim}0.1"$} & May
1997
\\
GJ~2069~BC    & 0.36"$\pm$0.01" & 0.45$\pm$0.3 & May 1997 \\
LHS~6158~AB & 0.62"$\pm$0.01" & 0.50$\pm$0.04 & Jan 1997 \\
Gl~381~AB & 0.18"$\pm$0.02" & 0.95$\pm$0.05 & Aug 1997 \\
Gl~487~AB     & 0.23"$\pm$0.01" & 0.7$\pm$0.1  & Feb 1997 \\
LHS~2887~AB  & 0.58"$\pm$0.01" & 0.15$\pm$0.03 & May 1997 \\ \hline
\end{tabular}
\label{tab_oa}
\end{table}

\section{Discussion}

% \subsection{Tidal circularisation}
%
%
%\begin{figure*}
%\psfig{height=12cm,file=eccentricite.ps,angle=-90}
%\caption{The Eccentricity versus Period diagram. The M dwarfs 
% circularisation limit is about 10 days.}
%\end{figure*}
%\label{eccentricity}
%
%{\em Apres examen l'interet d'une discussion de cette distribution
% parait modeste: la dependance en masse du temps de circularisation
% theorique est faible (de l'ordre de M^{1/3} semble-t'il), et il est
% donc normal que nous obtenions essentiellement la meme chose que
% pour les G. Il faudrait une statistique nettement meilleure pour
% mieux definir la periode limite, et pouvoir eventuellement raconter
% des choses interessantes. Eventuellement envisageable dans un article
% ulterieur elargissant cette discussion a un echantillon moins bien defini
% mais plus large de binaires M, car les biais potentiels sont ici plus
% ou moins negligeables.
% }

%\subsection{The solar neighbourhood sample}

After 2.5 years, we have found 12 new companions to the 127 M dwarfs
in our initial sample, and over the same period 6 additional
companions
% G41-14, G203-47, GJ1230, Reid & Gizis, LP476-207, LHS 1885, G89-32 
% (Henry et al.)
were discovered by others (we also independently found most of
those). While showing that the multiplicity information was previously
fairly incomplete, this should not be taken as the final number of
companions for this sample. Many stars still have few
radial-velocity measurements (the median is 5 % ?? ou 4?
but a number of stars only have 2 measurements), which span an
interval of at most 2.5 years. Our detectivity is thus very
significantly biased towards nearly equal-mass binaries (detectable as
double-lined binaries in a single measurement) or/and short periods. A
significant number of lower mass companions in wider orbits certainly
remains to be found in this sample.

The discovery of this large number of new multiple systems has two
opposite effects on estimates of the stellar density in the solar
neighbourhood. On the one hand, the new components within 9 pc
increase the previously underestimated local density, typically with
lower mass stars as they are necessarily fainter than their
primaries. On the other hand, a number of previously unknown M-dwarf
binaries are actually beyond the 9-pc distance limit of the sample, in
which they were initially included on the basis of an underestimated
photometric distance, reflecting the well known $\sim$2$^{3/2}$ volume
bias in photometric parallaxes of unrecognized binaries. This second
correction decreases the local density, more or less uniformly for all
masses. These two effects have been abundantly discussed in the
context of photometric luminosity functions, as a possible explanation
of their differences from the solar neighborhood luminosity function
(e.g. Kroupa 1995, and Reid \& Gizis 1997, for two contrasted views).
It is perhaps not always recognized that, beyond $\sim$5~pc, the solar
neighborhood luminosity function still has a significant photometric
component, and that it is thus also affected at some level by the same
two effects.

% Ou va-t'on cote echantillon?
% - 11 etoiles (10 systemes) sorties de 9pc: Gl49 (9.4pc), Gl70(11.2pc), 
%   LHS1885(11.4pc), Gl268.3(12.3pc), GJ2066(9.15pc), GJ2069A(12.8pc), 
%   GJ2069B(12.8pc), Gl381(12.3pc), Gl424(9.1pc), Gl487(10.2pc), 
%   LHS2887(15.8+-3.3pc, D_phot~11-12pc). Dont 2 gardee dans 9.25pc 
%   (GJ2066 et Gl424)
% - Par ailleurs 2 etoiles rentrent dans 9pc: Gl250B et Gl203.
% - au stade actuel, il reste 9 etoiles qui n'ont encore que des parallaxes 
%   photometriques: LP467-16, LP476-207 (triple SB2+speckle), LHS1723, G89-32
%   (binaire speckle, delta_K=2), LHS6158/G41-14 (triple, SB3(2O)+OA),
%   LHS2520 (M3.5, pi_phot=10.1pc dans Reid), GJ2097(M2V a d=25pc d'apres 
%   Henry et al., M1.5V a 32pc d'apres Reid), G165-8, LP229-17. 
%   3 etoiles (GJ2097  tres clairement, LP476-207 et LHS2520 tres 
%   probablement) sortent de 9pc. Combien d'autres? Mettre les parallaxes 
%   photometriques de Reid?
% A la louche, environ 10% de perte. Compense par 5 nouvelles
% composantes. Type spectral (ou M_V) moyen de ce qui rentre et de ce qui
% sort?
%

Taking together newly published trigonometric parallaxes (ESA, 1997;
Van Altena at al. 1995) and corrected photometric distances to the new
binaries, 13-M-dwarf systems listed within 9~pc in the preliminary
version of the
% as 14 stars in initial sample (GJ2069)
CNS3 are actually beyond this distance, while only one system (Gl~203)
enters this volume. The better observed 5.2-pc sample, on the other
hand, is essentially unaffected, with only the G041-014 system removed
($\sim$4.5~pc~$\rightarrow~\sim$8~pc) from it. The previously mentioned (e.g.
Henry et al. 1994) incompleteness of the nearby M-dwarf system sample 
beyond 5~pc is thus made more significant.
%(?? des nombres). 
% Pas tres evident a rendre quantitatif, vu les definitions d'echantillons 
% variees des uns et des autres: parallaxes trigo du Gliese uniquement
% ou distances spectroscopiques Reid+Gizis, plus divers degres de menage,
% plus selections de type spectral ou non, plus limite de declinaison
% variable... Mon impression actuelle est que tous ces braves gens ont
% un peu tendance a choisir la definition dont les fluctuations vont 
% ``dans le bon sens''.
%
Systems with bright secondaries are rapidly recognized as double-lined
binaries, and their inventory should thus be essentially complete. The
remaining 7 photometric parallaxes in this revised 9-pc sample can
therefore be considered as reliable. The 9-pc northern sample of systems
should thus now only significantly change through inclusion of
presently missing systems.
%The new binaries presented in this papers have for effect (1) to reject 7
% system out of 9 pc (confirmed by the new parallax measurements) 

The new multiple systems, on the other hand, add 6 stars to
the 9-pc sample of stars. 
% un petit point sur le nb total de binaires "connues" dans 9 pc, en
% eliminant les couples rejete et en ajoutant les compagnions??
Given the remaining selection biases in our binary search,
we feel that derivations of the binarity statistics and stellar
luminosity function within 9~pc would still be premature. 

Our sensitivity to lower companion masses and longer periods is
however quickly improving, so that this information will soon become
available. It will then be possible to more reliably estimate the
correction for unresolved binaries in photometric determinations of
the luminosity function, and hopefully settle the long-standing
controversy (e.g. Kroupa 1995 and Reid \& Gizis 1997) on its true
importance.

%dire quelque chose sur comment on complete le sample de reid
% Non. TF.
%dire clairement que pour l'instant on ne veut surtout pas faire de 
%statistique binarite parce que nous ne sommes pas complet....

% But now it is extremely delicate to estimate the luminosity
% function of the solar neighbourhood for $d>5pc$ because it is biased by the
% unknow binaries 
% Pas le seul probleme: il manque aussi des systemes, et ca, ca n'est pas
% nous qui allons les trouver. Sauf avec DENIS? Du coup je prefere ne pas
% en parler.

\begin{acknowledgements}
  We thank the technical staffs and telescope operators of OHP and CFHT
  for their support during these long-term observations. We are also
  grateful to Didier Queloz and Luc Weber for having developed the
  powerful data reduction package of the ELODIE spectrograph, and for
  their support in porting it to a different flavour of
  Unix. We are grateful to Gilles Chabrier, Isabelle Baraffe 
  and France Allard for useful discussions, and for communicating 
  unpublished results. 
  J.-L. B. would also like to thank Meghan Gray who reduced part of the
  AO data while being a student at CFHT in May-August 97.
  X.D. acknowledges support by the French Minist\`ere des
  Affaires \'Etrang\`eres through a ``Lavoisier'' grant for his 1~year stay at
  Observatoire de Gen\`eve.

  "This research has made use of the Simbad database,
  operated at CDS, Strasbourg, France"

\end{acknowledgements}

\end{document}